%% file: main.tex
\documentclass[]{pasj01}
\draft

\usepackage{xcolor}

\newcommand{\RCa}{black}

\begin{document} 
\Received{}
\Accepted{}

\input{definitions}

\input{s0_title}
\maketitle
\input{s0_abstract}
\input{s1_introduction}

\input{s2_data}

\input{s3_results}

\input{s4_discussion}

\input{s5_conclusions}

\newpage
\newpage

\input{s9_references}
\input{s9_figures}

\end{document}

%% file: definitions.tex
\def\Herschel{{\it Herschel}}
\def\Spitzer{{\it Spitzer}}
\def\WISE{{\it WISE}}
\def\AKARI{{\it AKARI}}

\def\um{$\mu \rm{m}$}

\def\Xco{$X_{\rm CO}$}
\def\Xunit{$\rm{cm}^{-2}$ $\rm{K}^{-1}$ $\rm{km}^{-1}$ $\rm{s}$}
\def\cmcm{$\rm{cm}^{-2}$}
\def\cmcmcm{$\rm{cm}^{-3}$}
\def\kms{$\rm{km}$ $\rm{s}^{-1}$}
\def\Kkms{$\rm{K}$ $\rm{km}$ $\rm{s}^{-1}$}
\def\vlsr{$v_{\rm LSR}$}
\def\degree{$^{\circ}$}
\def\Lsun{$L_{\solar}$}
\def\Msun{$M_{\solar}$}
\def\Msunyr{$M_{\solar}$ $\rm{yr}^{-1}$ }

\def\NH{$N(\rm{H_{2}})$}

\def\HII{H \emissiontype{II}}
\def\OIII{O \emissiontype{III}}
\def\SII{S \emissiontype{II}}

\def\Ha{H$\alpha$}

\def\lb{($l$, $b$)}
\def\lbeq{($l$, $b$)$=$}
\def\lbsim{($l$, $b$)$\sim$}
\def\radec{($\alpha_{\rm J2000}$, $\delta_{\rm J2000}$)}
\def\radeceq{($\alpha_{\rm J2000}$, $\delta_{\rm J2000}$)$=$}
\def\radecsim{($\alpha_{\rm J2000}$, $\delta_{\rm J2000}$)$\sim$}

\def\COa{\atom{C}{}{12}\atom{O}{}{}}
\def\COb{\atom{C}{}{13}\atom{O}{}{}}
\def\COc{\atom{C}{}{}\atom{O}{}{18}}

\def\Jeq{{\it J}$=$}
\def\Ja{{\it J}$=1$--$0$}
\def\Jb{{\it J}$=2$--$1$}
\def\Jc{{\it J}$=3$--$2$}

%% file: s0_title.tex
\title{FOREST unbiased Galactic plane imaging survey with the Nobeyama 45 m telescope (FUGIN): Possible evidence of cloud-cloud collisions triggering high-mass star formation in the giant molecular cloud M16 (Eagle Nebula)}


\author{Atsushi \textsc{nishimura}\altaffilmark{1*}}
\author{Shinji \textsc{fujita}\altaffilmark{2}}
\author{Mikito \textsc{kohno}\altaffilmark{2}}
\author{Daichi \textsc{tsutsumi}\altaffilmark{2}}
\author{Tetsuhiro \textsc{minamidani}\altaffilmark{3,4}}
\author{Kazufumi \textsc{torii}\altaffilmark{3,4}}
\author{Tomofumi \textsc{umemoto}\altaffilmark{3,4}}
\author{Mitsuhiro \textsc{matsuo}\altaffilmark{3}}
\author{Yuya \textsc{tsuda}\altaffilmark{6}}
\author{Mika \textsc{kuriki}\altaffilmark{7}}
\author{Nario \textsc{kuno}\altaffilmark{7,8}}
\author{Hidetoshi \textsc{sano}\altaffilmark{2,9}}
\author{Hiroaki \textsc{yamamoto}\altaffilmark{2}}
\author{Kengo \textsc{tachihara}\altaffilmark{2}}
\author{Yasuo \textsc{fukui}\altaffilmark{2,9}}

\altaffiltext{1}{Department of Physical Science, Graduate School of Science, Osaka Prefecture University, 1-1 Gakuen-cho, Naka-ku, Sakai, Osaka 599-8531, Japan}
\altaffiltext{2}{Department of Physics, Nagoya University, Furo-cho, Chikusa-ku, Nagoya, Aichi 464-8602, Japan}
\altaffiltext{3}{Nobeyama Radio Observatory, National Astronomical Observatory of Japan (NAOJ), National Institutes of Natural Sciences (NINS), 462-2 Nobeyama, Minamimaki, Minamisaku, Nagano 384-1305, Japan}
\altaffiltext{4}{Department of Astronomical Science, SOKENDAI (The Graduate University for Advanced Studies), 2-21-1 Osawa, Mitaka, Tokyo 181-8588, Japan}
\altaffiltext{5}{Institute of Space and Astronautical Science, Japan Aerospace Exploration Agency, Chuo-ku, Sagamihara 252-5210, Japan}
\altaffiltext{6}{Graduate School of Science and Engineering, Meisei University, 2-1-1 Hodokubo, Hino, Tokyo 191-0042, Japan}
\altaffiltext{7}{Department of Physics, University of Tsukuba, 1-1-1 Ten-nodai, Tsukuba, Ibaraki 305-8577, Japan}
\altaffiltext{8}{Center for Integrated Research in Fundamental Science and Technology (CiRfSE), University of Tsukuba, Tsukuba, Ibaraki 305-8571, Japan}
\altaffiltext{9}{Institute for Advanced Research, Nagoya University, Chikusa-ku, Nagoya, Aichi 464-8601, Japan}

\email{nishimura@p.s.osakafu-u.ac.jp}

\KeyWords{ISM: clouds --- ISM: individual objects (M16) --- stars: formation --- radio lines: ISM } 

%% file: s0_abstract.tex
\begin{abstract} 

M16, the Eagle Nebula, is an outstanding \HII \ region which exhibits extensive high-mass star formation and hosts remarkable "pillars". 
We herein obtained new \COa \ \Jeq 1--0 data for the region observed with NANTEN2, which were combined with the \COa \ \Jeq 1--0 data obtained using FOREST unbiased galactic plane imaging with Nobeyama 45-m telescope (FUGIN) survey. 
These observations revealed that a giant molecular cloud (GMC) of $\sim 1.3 \times 10^5$ \Msun \ is associated with M16, which is elongated by over 30 pc and is perpendicular to the galactic plane, at a distance of 1.8 kpc. 
This GMC can be divided into the northern (N) cloud, the eastern (E) filament, the southeast (SE) cloud, the southeast (SE) filament, and the southern (S) cloud.
We also found two velocity components (blue and red shifted component) in the N cloud. 
The blue-shifted component shows a ring-like structure, as well as the red-shifted component coincides with the intensity depression of the ring-like structure.
The position-velocity diagram of the components showed a V-shaped velocity feature.
The spatial and velocity structures of the cloud indicated that two different velocity components collided with each other at a relative velocity of 11.6 \kms.
The timescale of the collision was estimated to be $\sim 4 \times 10^5$ yr.
The collision event reasonably explains the formation of the O9V star ALS15348, as well as the shape of the Spitzer bubble N19.
A similar velocity structure was found in the SE cloud, which is associated with the O7.5V star HD168504.
In addition, the complementary distributions of the two velocity components found in the entire GMC suggested that the collision event occurred globally.
On the basis of the above results, we herein propose a hypothesis that the collision between the two components occurred sequentially over the last several $10^{6}$ yr and triggered the formation of O-type stars in the NGC6611 cluster. 

\end{abstract}

%% file: s1_introduction.tex
\section{Introduction}

High mass stars, such as O- and early B-type stars, contribute considerable amounts of energy to the galactic disk, and they dominate dynamics of the interstellar medium and influence galactic evolution. 
It is, therefore, crucial to understand the mechanisms of high-mass star formation in order to obtain a comprehensive understanding of galactic evolution. 
Considerable amounts of effort have been devoted toward this issue (e.g., for a review see \cite{zin07}).

O/early B stars ionize \HII \ regions, which are often associated with natal molecular gas. 
\HII \ regions are, therefore, a primary target for studying the mechanisms and initial conditions of high-mass star formation on the basis of observations of stars and molecular gas. 
One difficulty in tracing star formation in \HII \ regions is that ionization significantly affects natal gas, which is dispersed quickly after star formation. 
In order to minimize the influence of ionization, very young star formation needs to be investigated via molecular studies, and such samples need to be sufficiently large to yield statistically sound conclusions. 
Recent studies of giant molecular clouds (GMCs) near the Sun included Orion A \citep{nis15, fuk17b, tan19, ish19, nak19}, Orion B \citep{pet17, eno19b, fuj17, oha17}, Cygnus-X \citep{yam18, tak19, dob19}, M17 \citep{yam16, nis18, shi19}, M20 \citep{tor11, tor17}, Vela \citep{fuk16, san18, eno18}, and NGC6334/NGC6357 \citep{fuk18a}, are part of such efforts.
These \HII \ regions show very young star formation with an age of $\sim 10^5$ yr, providing us with hints regarding the initial cloud conditions and distributions prior to O star formation. 

The distance scale affected by ionization is given as $v_i \times$ age, where $v_i \sim 5$ \kms \ is the typical velocity of an ionization front (e.g., \cite{spi68}).
An age of $10^5$ yr yields $\sim 0.5$ pc as the distance, which is smaller than a typical star forming cloud (e.g., Orion and Vela) having a size of a few to $\sim 10$ pc. 
On the other hand, molecular gas within a 10 pc radius becomes significantly ionized if there are 10 or more O stars whose ages are $\sim$2 Myr, as in Westerlund2 \citep{fur09, oha10}, NGC3603 \citep{fuk14}, and [DBS2003]179 \citep{fuk20}. 
Natal gas directly related to star formation has already been lost in those regions, however, it is still possible that the rest of the gas outside 10 pc, which has yet to be ionized, may contain a hint of the initial conditions of the parental cloud and any early interactions with other clouds. 
It is noteworthy that recent large-scale mapping of GMCs has shown possible evidence of cloud-cloud collisions in molecular gas outside ionized regions, which suggests that collisions may trigger O/early B star formation in all of the \HII \ regions listed above, and perhaps many others \citep{fuk18b, oha18a, fuk18c, oha18b, hay18, koh18a, tor18a,  tor18b, koh18b, fuj19a, fuj19b, eno19a}.

M16, the Eagle Nebula, is one of the most well-known high-mass star formation sites (Figure \ref{fig:1}a; for a review, see \cite{oli08}).
At a distance of 1.8 kpc \citep{bon06, duf06, gua07}, it is located on the Sagittarius Arm, which consists of an open cluster (NGC6611), an \HII \ region known as Sh2-49 \citep{sha59}, Gum83 \citep{gum55}, RCW165 \citep{rod60} or W37 \citep{wes58}, some bright rimmed clouds (e.g., \cite{sug91}), and molecular clouds \citep{dam01}.
NGC6611 contains 52 known O/early-B stars \citep{eva05}, with the most massive one being the O4 star HD168076.
The age and total mass of NGC6611's stellar population have been estimated to be $1.3 \pm 0.3$ Myr \citep{bon06} and  $\sim 2.5 \times 10^{4}$ \Msun \ \citep{wol07}, respectively.
The molecular gas within $\sim 5$ pc from the cluster is heavily ionized, but a remnant of the parental gas still exists in spectacular pillars named elephant trunks (e.g., \cite{hes96, lev15}). 
Moreover, molecular clouds are present outside the ionization front created by the cluster \citep{hil12}, with active ongoing low-mass star formation in these clouds \citep{pov13}.
In addition, the association of two O-type stars, ALS15348 and HD168504, with the surrounding clouds suggests that high-mass star formation is also ongoing in the clouds.
The molecular clouds in M16 are, therefore, suitable to be used in investigating the initial conditions of cluster formation and the process of forming O-type and low-mass stars.

We carried out new CO \Jeq 1--0 observations with NANTEN2 toward M16 and combined them with already existing high-resolution FUGIN data. 
The combined data revealed the whole cloud and its surroundings at angular resolutions suitable for studies of star formation and cloud interactions with \HII \ regions. 
We herein present the first results of these observations. 
This paper is organized as follows: Section 2 provides descriptions of the CO observations, Section 3 presents the CO results, Section 4 presents a discussion of star formation in clouds, and Section 5 concludes the paper.

%% file: s2_data.tex
\section{Data set}

\subsection{\COa \ \Ja \ observations with the NANTEN2 telescope}

The entire extent of the molecular gas toward the M16 GMC was observed with the NANTEN2 4-m submillimeter telescope \citep{miz04} using \COa \ \Jeq 1--0 emission line at 115 GHz during the period from January 2012 to January 2013.
A superconductor-insulator-superconductor (SIS) mixer receiver was used for the front end, and a digital Fourier transform spectrometer with 16,384 channels of 1 GHz bandwidth, corresponding to velocity coverage and resolution of 2,600 \kms \ and 0.16 \kms, respectively, was used for the back end.
The double-sideband (DSB) system noise temperature was typically measured to be 110 K toward the zenith including the atmosphere during the observation period.
On-the-fly (OTF) mapping mode was used for observations, and each observation was carried out on a submap which has an area of 1 deg$^2$.
The pointing accuracy was checked every three hours and achieved within an offset of 25$''$.
The absolute intensity was calibrated by observing IRAS 16293-2422.
After initial data reductions, the final beam size (full width at half maximum, FWHM) was 180$''$, and the typical noise level was 0.42 K in the $T_{\rm mb}$ scale.

\subsection{FUGIN \COa, \COb, and \COc \ \Ja \ data}
\label{sec:data-fugin}

In order to investigate the detailed structures in the GMC, \COa, \COb, and \COc \ \Ja \ data sets obtained by the FUGIN survey with the Nobeyama 45-m telescope were analyzed.
The FUGIN project observed the galactic plane with a range of ($l = 10$\degree -- $50$\degree, $b \leq \pm1$\degree) using \COa, \COb, and \COc \ \Ja \ lines.
The FOur-beam REceiver System on the 45m Telescope (FOREST; \cite{min16}) and the FX-type correlator SAM45 \citep{kun11,kam12} were used to perform the observations.
OTF observations \citep{saw08} were carried out for each submap, which were divided into squares of $1 \times 1$ deg$^2$.
The angular resolution of the data sets was $\sim20''$, and the pointing accuracy was kept within 3$''$.
More detailed information on the observations and data reduction was provided by \citep{ume17}, and the data quality was discussed by \citep{tor19}.

The scan strategy of the FUGIN observations was optimized for large scale mapping.
In order to achieve this goal, an arbitrary ON position was scanned by any of the four beams of FOREST, and the observed data from each beam were merged to create fully sampled large scale maps.
Therefore, because of the different noise temperature in each beam, undesirable scanning effects appeared in some of the merged maps, especially when the weather conditions were less than ideal.
For the region containing M16, the map of $l=17$\degree--$18$\degree \ suffered from some scanning effects.
In order to remove these features, the CO data sets were smoothed to a 50$''$ resolution for this study.

In the current study, the CO data of $l=16.5$\degree--$17.5$\degree \ and $b=0$\degree--$1$\degree \ were cut out and used.
It is worth noting that although the distribution of the M16 GMC about the galactic latitude ($b = 0$\degree -- $1.2$\degree) exceeds the survey area of the FUGIN ($b \leq \pm1$\degree), the FUGIN data sets are still useful for this study as the majority of the molecular gas emission is distributed within the area mapped by FUGIN.

%% file: s3_results.tex
\section{Results}

\subsection{Large scale molecular distribution with NANTEN2}

Figure \ref{fig:1}b shows the large-scale integrated intensity distribution of the \COa \ \Jeq 1--0 emission in a 3.5\degree $\times$ 2\degree \ region containing the M16 and M17 GMCs obtained with the NANTEN2 telescope.
Intensities were calculated by integrating the spectra between 0 and 40 \kms, which almost corresponds to the velocity range of near side of the Sagittarius Arm.
The maximum intensity of \COa \ \Jeq 1--0 was found at the position of the M17 GMC ($l$=15.001\degree, $b$=$-$0.667\degree) with an intensity of 382 \Kkms, and the second maximum intensity was found at the position of the M16 GMC ($l$=17.035\degree, $b$=0.867\degree) with an intensity of 165 \Kkms.
Both GMCs are located at the edge of a molecular super shell \citep{mor02, gua10, com18}. 
The M17 GMC is distributed almost parallel to the galactic plane, whereas the M16 GMC is distributed almost perpendicular to the plane.
The size of the M16 GMC is 30 $\times$ 10 pc.
The intensity peak of \COa \ \Jeq 1--0 and cluster NGC6611 are located at the northwestern side of the GMC.
The total mass of the GMC is estimated to be $1.3 \times 10^5$ \Msun \ from \COa \ \Jeq 1--0 transition using an \Xco \ factor of $1.8 \times 10^{20}$ \Xunit \ \citep{dam01}.

Figure \ref{fig:2} shows a latitude-velocity diagram of the M16 GMC.
Most of the emission was found in the velocity range 15--28 \kms.
The GMC has a uniform velocity distribution in regions below $b<0.7$\degree \ with an intensity peak at 24 \kms , whereas it has a double-peaked velocity at 20 \kms \ and 22--26 \kms \ for regions above $b>0.7$\degree.
A diffuse component was found for regions below $b<0.6$\degree, which had a velocity range of 30--40 \kms.
This diffuse component seems to be unrelated to the GMC; therefore we used the data with a velocity range of 10--30 \kms \ for the analyses of the GMC hereafter.

\subsection{Detailed molecular distribution with FUGIN}

Figure \ref{fig:3} shows a velocity-integrated intensity map of \COa \ \Jeq 1--0 obtained by FUGIN.
Data were smoothed to a 50$''$ resolution to remove the scanning effects (see Section \ref{sec:data-fugin}).
Intensities were calculated by integrating the spectra between 10 and 32 \kms.
NANTEN data (180$''$ resolution) were used to show the entire distribution of the GMC, which extend up to $b=1.2$\degree \ (and hence exceed the FUGIN coverage beyond $b=1$\degree).
The GMC can be divided into five subregions: a northern (N) cloud, an eastern (E) filament, a southeastern (SE) cloud, a southeastern (SE) filament, and a southern (S) cloud.
The locations and definitions of boundaries of these subregions are shown in Figure \ref{fig:3}.
The N cloud is the most massive cloud in the GMC, which is distributed around $l>$ 0.7\degree \ and is associated with active star formation regions, including the NGC6611 cluster and the Spitzer bubble N19, the latter of which is associated with the O-type star ALS15348.
The cloud shows many short filamentary structures with CO emission, and the same structures are also seen in dust observations made with Herschel \citep{hil12}.
The E filament connects the N cloud and the SE cloud and is elongated from \lbeq (17.05\degree, 0.4\degree) to \lbeq (17.1\degree, 0.65\degree).
The width and length of the filament are 2 pc and 6 pc, respectively.
The SE cloud has a \textcolor{\RCa}{rounded} shape and is located at $b<$0.4\degree, which is associated with a filament-shaped structure, the SE filament, located at $l<$16.83\degree \ and $b<$0.2\degree.
The O-type star HD168504 is located at almost the center of the cloud.
The velocity distributions of the SE cloud and the SE filament are different, that is, 19--27 \kms \ for the SE cloud and $\sim$15 \kms \ for the SE filament (see also Figure \ref{fig:4}).
The S cloud is located at the southern side of the NGC6611 cluster.

Velocity distributions are very complicated in the GMC, as seen in the velocity channel maps shown in Figure \ref{fig:4}.
The GMC consists of many small filaments and cores, whose spatial distributions are drastically different in each velocity range.
The N cloud mainly consists of a blue-shifted ring-like structure (\vlsr $=$16--20 \kms) and a red-shifted cloud (\vlsr $=$24--28 \kms).
For the E filament, intense well-defined filaments are found in the red-shifted part (\vlsr $=$23--26 \kms), whereas the blue-shifted part  (\vlsr $=$18--22 \kms) shows a more diffuse distribution.
The blue-shifted component of the SE cloud (\vlsr $=$16--22 \kms)  consists of a diffuse component showing no internal structure, whereas the red-shifted component of the cloud (\vlsr $=$24--28 \kms) shows a filamentary structure.
The velocity ranges of the blue- and red-shifted components are slightly different for the N cloud, E filament, and SW cloud, whereas all regions can be divided into two velocity components using a central velocity of $\sim 22$ \kms.
The spatial distributions are also significantly different between the various components.

\subsection{Detailed velocity distributions of the N cloud}

Figure \ref{fig:5} shows the distribution of the blue-shifted component (\vlsr $< 20$ \kms) toward the N cloud.
This component shows a ring-like structure centered at \lbeq (17.08\degree, 0.93\degree) with a diameter of $\sim$5 pc (Figure \ref{fig:5}a). 
The intensity depression of the molecular gas clearly coincides with the ring-like distribution of the 8 $\mu$m Spitzer image (Figure \ref{fig:5}b) and also with the distribution of the \HII \ region (Figure \ref{fig:5}c).
The O9V star ALS15348 ($l=$17.083\degree, $b=$0.974\degree), which is an ionizing star in the Spitzer bubble N19, is located at the center of the molecular ring with an offset of $\sim$1 pc toward the NW direction.
The width of the molecular ring is broader on the northern side (1--5 pc) than on the southern side ($\sim$0.5 pc), the latter of which adjoining the NGC6611 cluster, indicating that the southern side of the component became ionized and dissociated by feedback from the cluster.
Figure \ref{fig:6} shows the distribution of the red-shifted component (\vlsr$> 24$ \kms), which has no clear correspondence with the distribution of the N18 bubble.
The intensity peaks near the contact point to the cluster ($l=$17.038\degree, $b=$0.876\degree) and decreases precipitously toward the cluster, indicating the component is also affected by feedback from the cluster.

Figure \ref{fig:7} shows a longitude-velocity diagram of the Spitzer bubble N19.
The NW side of the bubble ($b=$0.98\degree--1.05\degree; Figure \ref{fig:7}a) mainly consists of a blue-shifted component, which seems to have no velocity gradient.
A red-shifted component is only found around $l=$17.0\degree, which is connected to the blue-shifted component with the intermediate-velocity emission.
In the central part of the bubble ($b=$0.92\degree--0.98\degree; Figure \ref{fig:7}b), most of the molecular gas is found in the intermediate-velocity range between the blue- and red-shifted components.
The velocity width is broader within the bubble compared to the outer part of the bubble.
The SE side of the bubble ($b=$0.85\degree--0.92\degree; Figure \ref{fig:7}c) shows a complicated velocity distribution, which can be divided into three parts: a blue-shifted component, a red-shifted component, and an intermediate bridging component.
The velocity bridge features are found at $l=$17.03\degree, 17.09\degree, \ and 17.2\degree.
The velocity broadening found at the interface to the cluster around $l=$17.00\degree--17.04\degree \ indicates that feedback from the cluster likely accelerated the blue-shifted component to 15 \kms \ and the red-shifted component to 30 \kms \ at the surface layer of the cloud.

The spatial relationship between the blue- and red-shifted components is shown in Figure \ref{fig:8}.
The distribution of the red-shifted component clearly corresponds to the distribution of the intensity depression of the blue-shifted component.
In addition, the intensity of the DSS Red (which roughly corresponds to the H$\alpha$ emission) is detected in the area where the intensities of both velocity components are depressed.
The complementary spatial distributions of the two velocity components and the (proxy) H$\alpha$ emission are found for the entire region of the N cloud with a spatial scale of $\sim 10$ pc.
Figure \ref{fig:9} shows a position-velocity diagram \textcolor{\RCa}{with respect to the O star ALS15348}.
The integration area and directions are represented by the green rectangle in Figure \ref{fig:8}.
The velocity distribution shows a V-shape with a maximum velocity of 30 \kms \ at a distance of 2 pc from ALS15348 at \lbeq (17.038\degree, 0.900\degree).
The gradient of the intensity peak velocity is steeper toward the cluster side than on the ALS15348 side.

\subsection{Large scale velocity distribution}

Figure \ref{fig:10} shows an intensity-weighted mean velocity map, with contours showing the integrated \COa \ \Jeq 1--0 emission intensity for reference of the distribution of the GMC.
On a large scale, the GMC is surrounded by diffuse molecular gas \textcolor{\RCa}{which mainly consists of the red-shifted component for the lower longitude and the blue-shifted component for the higher longitude.
The two velocity components are merged on the GMC.
In the smaller scale, some velocity gradients are found on the GMC, whereas the velocity gradient is not clearly seen in the larger scale.
}
In the M16 GMC, the two velocity components are jumbled and their spatial features seem to be complicated.
The GMC consists of blue- and red-shifted components that are connected by an intermediate-velocity component.
For the N cloud, the red-shifted component is surrounded by a blue-shifted component, which is consistent with their distributions shown in Figure \ref{fig:8}.
The E filament and SE cloud also consist of two velocity components clearly divided by an intermediate-velocity layer, which suggests that they have complementary distributions.
The blue- and red-shifted components in the SE cloud show complementary distributions, as shown in Figure \ref{fig:10-1}.
The O-type star HD168504 ($l=$17.026\degree, $b=$0.345\degree) is located at the boundary between the two velocity components.
The 24 $\mu$m emission, indicating the warm dust mainly heated by the \HII \ region G017.037+00.320 \citep{and14}, is associated with HD168504 and is distributed in the area where the CO intensities of both velocity components are depressed.
The spatial distributions of the two velocity components and the \HII \ region are similar to that of the N cloud.

Figure \ref{fig:11} shows the spatial distributions of the blue- and red-shifted components of the GMC.
The complementary distributions of the N cloud and the SE cloud are also clearly shown in the figure.
In addition, complementary distributions can be seen for the E filament, the S cloud, and the SE filament.
The SE filament mainly consists of a single velocity component of \vlsr$\sim$15 \kms, whereas the red-shifted component of the SE cloud surrounds the filament; thus, their spatial relationships are complementary.
The spatial relationships of the two velocity components in the E filament and the S cloud are more complicated than in the other subregions, although there are some complementary distributions between them.
The two velocity components in the E filament are not clearly separated like in the SE cloud, whereas the intensity peaks of the red-shifted component are located in the intensity depression of the blue-shifted component, and vice versa.
The intensity distribution of the S cloud is weaker and more diffuse compared to the E filament, whereas the trends of the two velocity components are similar to those of the E filament.
Furthermore, complementarity is also found across the entire scale of the GMC.
The red-shifted component of the GMC is surrounded by blue-shifted gas, and the area where the intensity of the red-shifted component is depressed within the GMC is filled by a blue-shifted component.
The blue-shifted component in a subregion, which is bounded by the E filament, the SE cloud, and S cloud, also shows complementary distributions with the red-shifted components in the surrounding subregions.

%% file: s4_discussion.tex
\section{Discussion}
\subsection{Star formation in the Spitzer bubble N19}

In the present study, a careful inspection of the velocity channel distributions of the \COa \ \Jeq 1--0 transition revealed that the M16 GMC can be divided into two velocity components, that is, a blue-shifted component (\vlsr $=$9.2--19.6 \kms) and a red-shifted component (\vlsr$=$24.2--31.3 \kms), which have significantly different spatial distributions.
In particular, the N cloud shows very clear complementary distributions between the two velocity components, whereby the smaller red-shifted cloud is surrounded by a larger blue-shifted cloud with a definitive boundary between them (Figure \ref{fig:8}).
The complementary distribution of different velocity channels is a signature of colliding clouds (\cite{fur09, oha10, shi13, dob14, fuk17b}).
Numerical simulations have shown that a supersonic collision of different-sized molecular clouds makes a hole behind the collision interface of the larger cloud, where gas from both clouds merges and becomes strongly compressed \citep{hab92, tak14}.
This hole in the larger cloud coincides with the path of the smaller cloud; therefore synthetic observations of this event shows a complementary distribution between the two velocity components corresponding to the larger and smaller clouds \citep{tor17, fuk18b}.
\textcolor{\RCa}{
Furthermore, for the case of that the collision direction is separated from the line of sight (LOS), the larger cloud is expected to be observed as a ring-like shape opening at the direction toward the approaching path of the smaller cloud \citep{tor15}.
For the N cloud, the blue-shifted component shows a ring-like structure opening at the Eastern side, and its opening region is filled with the red-shifted component (Figure \ref{fig:8}).
This geometric relation suggests that the collision direction is toward an east-west direction.
}
The center velocities of the blue- and red-shifted components of the N cloud were calculated to be 18.0 and 26.2 \kms, respectively, and the velocity separation along the LOS of the two velocity components was 8.2 \kms.
Therefore, it is a supersonic collision if the components are colliding.
The V-shaped velocity feature found in the N cloud (Figure \ref{fig:9}) further supports the collision scenario, as studies on numerical simulations have shown that this type of feature is observed in the late phase of collisional interactions of different-sized clouds \citep{tak14, haw15a, haw15b, fuk18b}.
All the results of the detailed analyses of the velocity structure of the molecular gas are consistent with the theoretical findings and indicate that the blue- and red-shifted components are the result of collisional interactions.
The collision velocity is estimated to be 11.6 \kms, if we assume a collision angle of 45\degree \ toward the LOS.
The timescale of the collision for the N cloud is roughly estimated to be $\sim 4 \times 10^{5}$ yr as the cloud size ($\sim 5$ pc) divided by the collision velocity.
\textcolor{\RCa}{
In general, two of clouds moving on the same direction toward the LOS is rare, therefore considering the some collision direction toward the LOS (typically 45\degree) is reasonable assumption.
However, the colliding clouds in the region having same traveling direction is possible.
If in the case, collision time scale is $\sim30$\% greater than the estimated value above.
}
The mass of the molecular gas and the peak column density for the blue- and red-shifted components are calculated to be $2.2 \times 10^4$ \Msun \ and $2.4 \times 10^4$ \Msun, and $2.3 \times 10^{22}$  \cmcm \ and $2.8 \times 10^{22}$ \cmcm, respectively, assuming an X-factor of 1.8 \Xunit \ \citep{dam01}.

The isolated O9V star ALS15348, an ionizing star of the Spitzer bubble N19, is surrounded by an \HII \ region with a radius of $\sim 1$ pc.
The formation timescale of the \HII \ region is roughly estimated to be $\sim 2 \times 10^5$ yr by assuming a typical \textcolor{\RCa}{velocity of the ionization front} of $\sim 5$ \kms \ \citep{hos05, fuk17b}.
Therefore, the molecular material surrounding the young \HII \ region may have retained some of the initial material and physical conditions that were present at the time of star formation; such a notion is worthy of investigation. 
The star is located on the border of the two velocity components at $l=$17.083\degree \ and $b=$0.974\degree \ (Figures \ref{fig:8} and \ref{fig:11}), indicating that the star was possibly formed by collisional interactions.
As the velocity separation of the two components is high ($\sim$8 \kms), the collision induced a high mass accretion rate in the range of up to 10$^{-3}$ to 10$^{-4}$ \Msunyr \citep{ino13, ino18}.
By assuming this high mass accretion rate, ALS15348 may have formed during an estimated collision time scale of $\sim 4 \times 10^{5}$ yr; if true, this timescale can be self-consistently explained by the collision scenario.
\textcolor{\RCa}{
The Spitzer bubble N19 is classified as a "complete or closed ring" in \citet{chu06}.
The Collect \& Collapse model (e.g., \cite{elm77}) is usually discussed as the origin of the Spitzer bubbles (e.g., \cite{deh10}).
In the model, an expanding \HII \ region sweeps up ambient material; therefore the model predicts that the ambient material possess a particular kind of velocity pattern \citep{tor15}.
However, the surrounding CO gas of the Spitzer bubble N19 shows no expanding motion (Figure \ref{fig:7}).
}
We here propose a collision scenario to explain the formation of ALS15348 and Spitzer bubble N19 following the scheme proposed for other Spitzer bubble RCW120 \citep{tor15} and RCW79 \citep{oha18a}.
Figure \ref{fig:12} shows a schematic diagram of the collision scenario for the N cloud.
Figures \ref{fig:12}a--c represent the projected spatial distributions of the two velocity components before collision and during collision as well as their current states after the collision.
We assumed that the blue-shifted component collided with the red-shifted component from the direction of higher galactic latitude and farther away along the LOS.
We also assumed that the NGC6611 cluster is associated with the red-shifted component as the \vlsr \ value of the cluster member, typically $\sim30$ \kms \citep{xu18}, is similar to that of the red-shifted component.
The age of the NGC6611 cluster is estimated to be several Myr \citep{hil93, bel00, bon06}; therefore, it is considered that the red-shifted component interacted with, and was compressed by, feedback from the cluster before the blue-shifted component started to collide with it (Figure \ref{fig:12}a).
After contact was made between the two velocity components (Figure \ref{fig:12}b), the interface layer was likely compressed strongly, as intimated by the scheme proposed by \citet{tor15}, which was seen in other numerical simulations \citep{hab92, ana09a, ana09b, ana10, ino13, tak14, mat15, wu15, wu17a, ino18, wu18}.
An O9 type star that formed in $\sim 10^5$ yr in the shocked layer likely ionized the two surrounding velocity components by ultraviolet (UV) feedback with a few $10^5$ yr (Figure \ref{fig:12}c).
The \HII \ region eventually formed between the two velocity components, as currently observed.

The UV feedback from the O type stars in the NGC6611 cluster and ALS15348 is another possible explanation for the distribution of the two velocity components, as the N cloud is obviously affected by strong feedback (see Figure \ref{fig:5}b and \ref{fig:6}b).
If the cloud was accelerated by feedback, a velocity shift as a function of distance from the stars should exist.
However, the intensity-weighted mean velocity map shows no clear velocity gradient centered at the O type stars (Figure \ref{fig:10}).
Furthermore, it is highly unlikely that UV feedback accelerated the different components to create the large-scale complementary distribution ($\sim 5$ pc) as observed.
Therefore, UV feedback does not seem to adequately explain the existence of the two velocity components in the N cloud.

\subsection{Star formation in the SE cloud}

In addition to the N cloud, complementary distributions were also found in the SE cloud (Figures \ref{fig:10} and \ref{fig:11}).
In this region, the two velocity components do not surround each other, as seen for the N cloud, but rather they are distributed mutually exclusively and are connected only on one side of the component boundary.
Synthetic observations from numerical simulations of collisions of similar-sized clouds showed similar types of complementary distributions, whereby a compressed layer formed as a plane between two clouds, and the clouds were observed as two velocity components bordered by the shocked layer \citep{wu17a, bis17}.
It is, therefore, possible that the two velocity components in the SE cloud are the result of colliding clouds.
If this is true, then the collision velocities are likely to be similar to those in the N cloud.

The O7.5V star HD168504 is isolated from other O-type stars in M16, and its similar distance, radial velocity, and proper motion with the NGC6611 cluster \citep{xu18} suggest that it is associated with the M16 GMC.
HD168504 is located at the center of the SE cloud in the integrated intensity map (Figure \ref{fig:3}), which is also along the border of two velocity components at $l=17.026$\degree \ and $b=0.345$\degree \ (Figure \ref{fig:10-1}).
Its position indicates that the star's formation was possibly triggered by the collision of the two velocity components.
As the velocity separation of the two components is high ($\sim$8 \kms), the collision was able to enhance the mass accretion rate enough to form HD168504; this scenario was also proposed to explain the formation of the O9V star ALS15348 (see Section 4.1).
The spatial relationships among the two velocity components and the \HII \ region around HD168504 are also very similar to the case of the N cloud.
The blue- and red-shifted components are complementarily distributed, and the cavity of the two velocity components corresponds to the distribution of the \HII \ region associated with the star.
We, therefore, conclude that HD168504 possibly formed by the same collision event as ALS15348.

Another possibility that explains the existence of the two velocity components is UV feedback from the NGC6611 cluster.
Figure \ref{fig:13} shows the spatial distribution of the 8 $\mu$m emission and the two velocity components of the CO emission.
The cluster side of the red-shifted component of the E filament and the SE cloud are associated in the 8 $\mu$m emission, indicating that the components interacted with, and are affected by UV emission from the cluster.
A temperature decrease was found in the E filament as a function of distance from the cluster.
The temperature peaks at the boundary of the cluster side of the filament and decreases with increasing distance from the cluster (see panel 23.0 \kms \ of Figure \ref{fig:4}).
The SE cloud shows no clear temperature gradient along the direction to the cluster.
These trends of temperature distribution of the molecular gas are similar to the observations of dust temperature made by Hershel \citep{hil12}. 
In addition, the velocity gradients along the direction to the cluster are not found in the E filament or the SE cloud.
We, therefore, conclude here that while the UV emission from the cluster is heating the E filament, it does not accelerate the E filament or the SE cloud, and hence it cannot be the cause of the two observed velocity components.

\textcolor{\RCa}{
\subsection{Possibility of the formation of the NGC6611 cluster by a cloud collision}
}


\textcolor{\RCa}{
The NGC6611 cluster is a more active and older system than the regions discussed above; the cluster contains 13 O type stars \citep{eva05}, and its age is estimated to be several Myr \citep{hil93, bel00, bon06}.
The molecular gas surrounding the cluster was heavily dissociated and dispersed by UV feedback from the cluster, and hence it is difficult to find any initial conditions of the cluster formation from the molecular remnant directly.
However, the complementarity of the spatial distributions of the two velocity components and the \HII \ region around the cluster is similar with that found in the N and SE clouds (Figure \ref{fig:15}), implying that the similar mechanism might originate the cluster.
The complementary distribution of molecular gas found in the S cloud and the SE filament and indeed across the entire scale of the GMC (see Section 3.4, Figure \ref{fig:11}) suggests that the GMC possibly experienced a cloud-cloud collision at global scales.
It is, therefore, still possible that the cluster was formed by a cloud collision that occurred prior to the current collision by several million years.
The GMC is located in the Sagittarius Arm, which also contains several CO clouds along the galactic plane around M16 and M17 (Figure \ref{fig:1}b).
M16 and M17 may potentially have arisen from multiple natal-cloud collisions because the high number density of molecular clouds favors more frequent collisions than the average in the galactic disk.
Furthermore, the M16 GMC is located at the boundary of a molecular shell (\cite{mor02, gua10, com18}; see also Figure \ref{fig:1}b) that might have formed by past feedback from ancient OB-type stars and supernova explosions, indicating that the GMC has possibly interacted with the shell.
If the M16 GMC has originated from the interaction between galactic atomic/molecular clouds and the shell, the GMC might have experienced multiple events.
}

%% file: s5_conclusions.tex
\section{Conclusions}

We analyzed large-scale CO distributions of the region of M16, the Eagle Nebula, taken with NANTEN2 and the FUGIN project.
A GMC of $1.3 \times 10^5$ \Msun \ is associated with M16, which is elongated perpendicularly to the galactic plane over 35 pc at a distance of 1.8 kpc. 
The cloud consists of two velocity components, that show complementary spatial distributions, suggesting dynamical interactions between them.
We, hence, summarize the present work as follows:

\begin{enumerate}
\item
The M16 GMC has a size of 10 pc $\times$ 30 pc in $l$ and $b$, and shows two velocity components of 9.2--19.6 \kms \ (blue-shifted component) and 24.2--31.3 \kms \ (red-shifted component). 
The total mass of the GMC is estimated to be $1.3 \times 10^5$ \Msun determined from the \COa \ \Jeq 1--0 transition observations, using an \Xco \ factor of $1.8 \times 10^{20}$ \Xunit.
The GMC mainly consists of the N cloud, E filament, SE cloud, SE filament and S cloud.

\item
The GMC is associated with 13 known O-type stars most of which are within the NGC6611 cluster.
We saw a clear intensity depression of molecular gas toward the O stars, which is likely due to photoionization: one is toward the heart of the Eagle Nebula at ($l$,$b=$16.95\degree, 0.85\degree) and the other is toward the Spitzer bubble N19 ($l$,$b=$17.06\degree, 1.0\degree). 

\item
For the N cloud, the blue-shifted component shows a ring-like structure, and the distribution of the red-shifted component coincides with the intensity depression of the blue-shifted component.
This complementary distribution, as well as their different velocities, suggests that larger and smaller clouds are colliding.
The V-shaped velocity feature found in the N cloud also supports this collision.
The collision velocity and timescale were estimated to be 11.6 \kms \ and $\sim 4 \times 10^5$ yr, respectively.

\item
The isolated O9V star ALS15348, an ionizing star of the Spitzer bubble N19, is located at the boundary of the two velocity components, which indicates that it possibly formed by compression arising from the collision between the components.
The star is surrounded by an \HII \ region with a size of $\sim 1$ pc, whose distribution coincides with the intensity depression of the two velocity components.
The collision scenario consistently explains the geometry of the molecular gas, O type star, and \HII \ region and the timescale of O type star formation therein.

\item
For the SE cloud, the blue- and red-shifted components show complementary distributions, indicating that similar-sized clouds are colliding.
The isolated O7.5V star HD168504 is located at the boundary of the two velocity components, and the spatial relationship among the molecular cloud, \HII \ region, and O-type star is similar to those of ALS15348, indicating that HD168504 may have possibly formed during the same process that resulted in ALS15348.

\item
Cluster NGC6611, which contains older stellar members aged several $10^6$ yr, ionized the surrounding molecular material, and it is therefore difficult to investigate the initial conditions of star formation.
However, the existence of complementary large-scale distributions implies that an older cloud collision may have occurred several $10^6$ yr ago.
Because the M16 GMC is located in the Sagittarius Arm and at the boundary of a molecular shell, the frequency of cloud collisions could theoretically be enhanced.
Therefore, if we assume that the GMC experienced multiple collision events within several $10^6$ yr, it is possible that the NGC6611 cluster formed during an older collision event.

\end{enumerate}

\begin{ack}

This work was supported by JSPS KAKENHI grant numbers,
18K13580
.
The authors would like to thank the members of the NANTEN2 group for support on the observation and telescope operation. 
The authors would like to thank the members of the 45-m group of Nobeyama Radio Observatory for support during the observation. 
The Nobeyama 45-m radio telescope is operated by Nobeyama Radio Observatory, a branch of National Astronomical Observatory of Japan.
Data analysis was carried out on the open use data analysis computer system at the Astronomy Data Center, ADC, of the National Astronomical Observatory of Japan.
This research made use of astropy, a community-developed core Python package for Astronomy \citep{ast13}, in addition to NumPy and SciPy \citep{wal11}, Matplotlib \citep{hun07} and IPython\citep{per07}.
This research has made use of the VizieR catalogue access tool, CDS, Strasbourg, France. The original description of the VizieR service was published in \citet{och00}.

\end{ack}

%% file: s9_figures.tex
\newpage

\begin{figure}[t]
 \begin{center}
  \includegraphics[width=\textwidth]{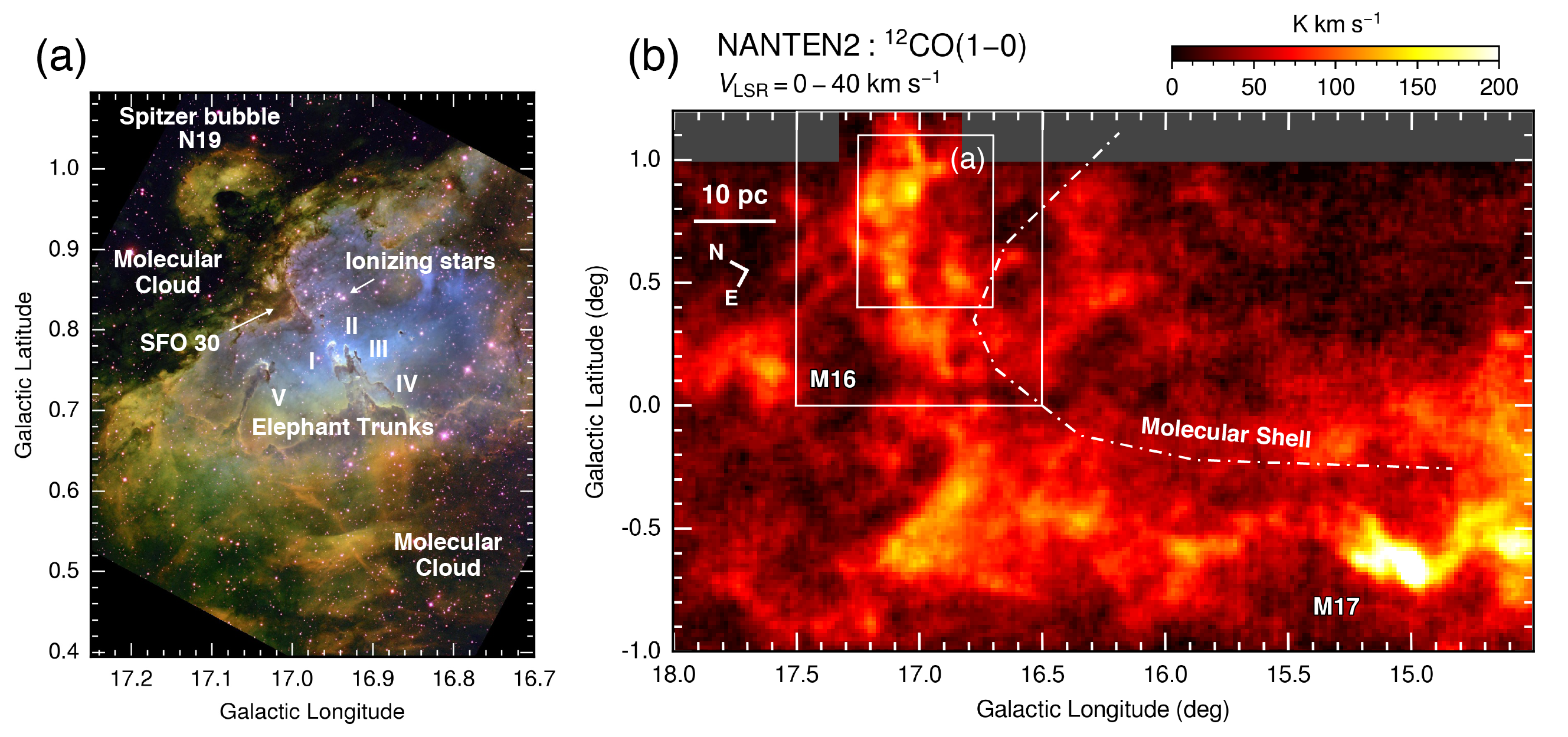}
 \end{center}
\caption{
(a) Optical composite image of the Eagle Nebula taken with the 0.9 m telescope at the Kitt Peak Observatory with the NOAO Mosaic CCD camera (Credit: T.A.Rector, B.A.Wolpa and NOAO).
Green, blue and red show H$\alpha$, [\OIII] and [\SII]. 
The pillars in Elephant Trunks (I--V), the Spitzer bubble N19 \citep{chu06}, the bright rimed cloud SFO30 \citep{sug91} and other objects are labeled in the figure.
(b) Integrated intensity map of \COa \ \Jeq 1--0 distribution toward the Sagittarius-Carina Arm including M16 and M17 GMCs taken with the NANTEN2 telescope. 
The intensity is integrated over 0 to 40 \kms.
The area shown in Figure \ref{fig:1}a and the definition of the boundary of the M16 GMC used the study are shown as white boxes with labels.
The limit of the molecular shell \citep{gua10} is shown by white dot-dashed curve.
}
\label{fig:1}
\end{figure}

\begin{figure}[t]
 \begin{center}
  \includegraphics[width=0.6\textwidth]{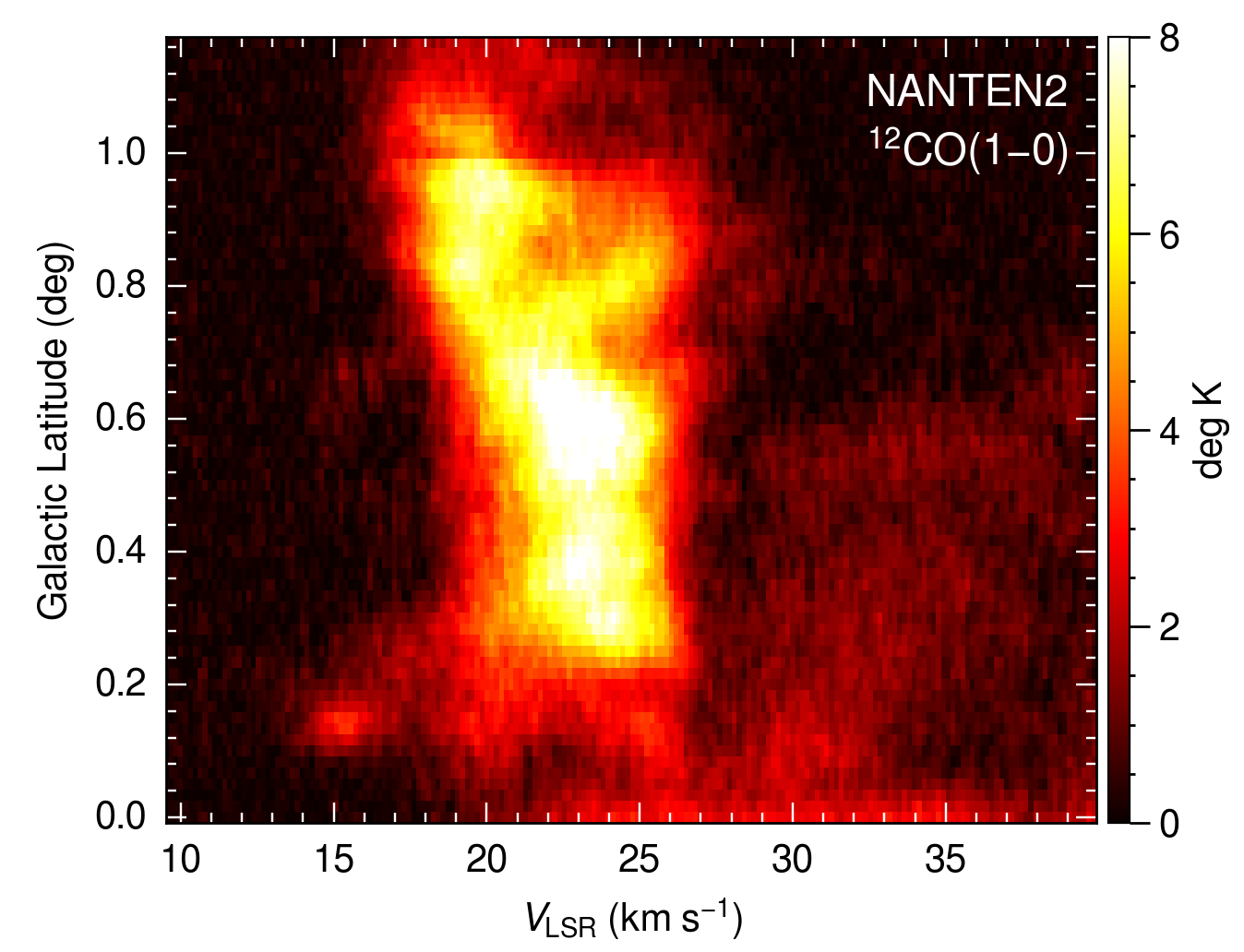} 
 \end{center}
\caption{
Latitude-velocity ($b$-$v$) diagram of the M16 GMC for the emission of \COa \ \Jeq 1--0 taken with the NANTEN2 telescope.
The intensity is integrated over $l=$17.5\degree \ and 16.5\degree.
}
\label{fig:2}
\end{figure}

\begin{figure}[t]
 \begin{center}
  \includegraphics[width=0.8\textwidth]{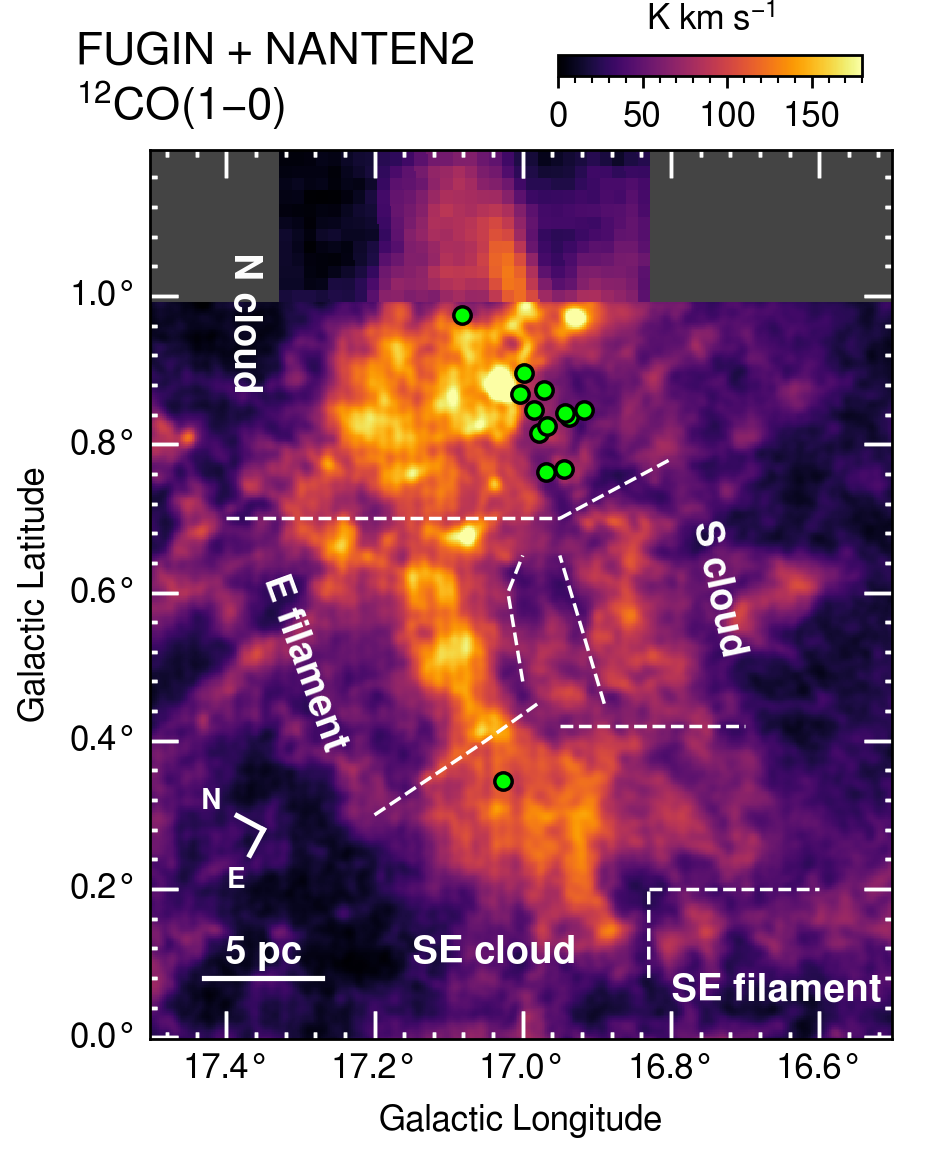}
 \end{center}
\caption{
Integrated intensity maps of \COa \ \Jeq 1--0 with a peak intensity of 276.2 \Kkms \ toward the M16 molecular clouds obtained by FUGIN project and NANTEN2 telescope.
The spectra with velocity range of 10 \kms $<$ \vlsr $<$ 32 \kms \ are integrated.
O-type stars \citep{eva05} are shown as green circles.
}
\label{fig:3}
\end{figure}

\begin{figure}[t]
 \begin{center}
  \includegraphics[width=\textwidth]{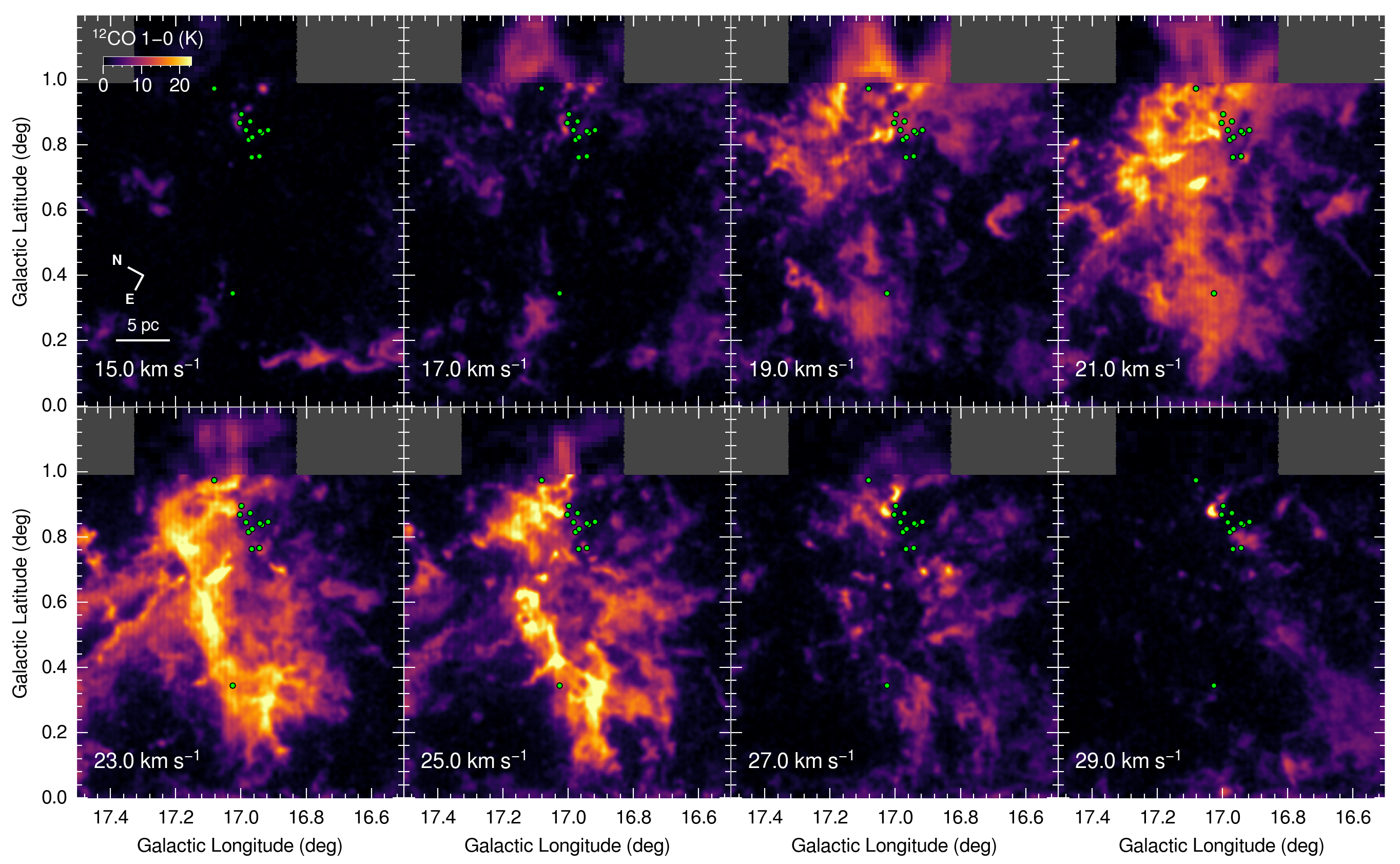}
 \end{center}
\caption{
Velocity channel maps of \COa \ \Jeq 1--0 FUGIN and NANTEN2 data for the velocity range 14 \kms $<$ \vlsr \ $<$ 30 \kms \ made every 2.0 \kms.
The center velocity is shown in the bottom left corner of each panel.
O-type stars \citep{eva05} are shown as green circles.
}
\label{fig:4}
\end{figure}

\begin{figure}[t]
 \begin{center}
  \includegraphics[width=\textwidth]{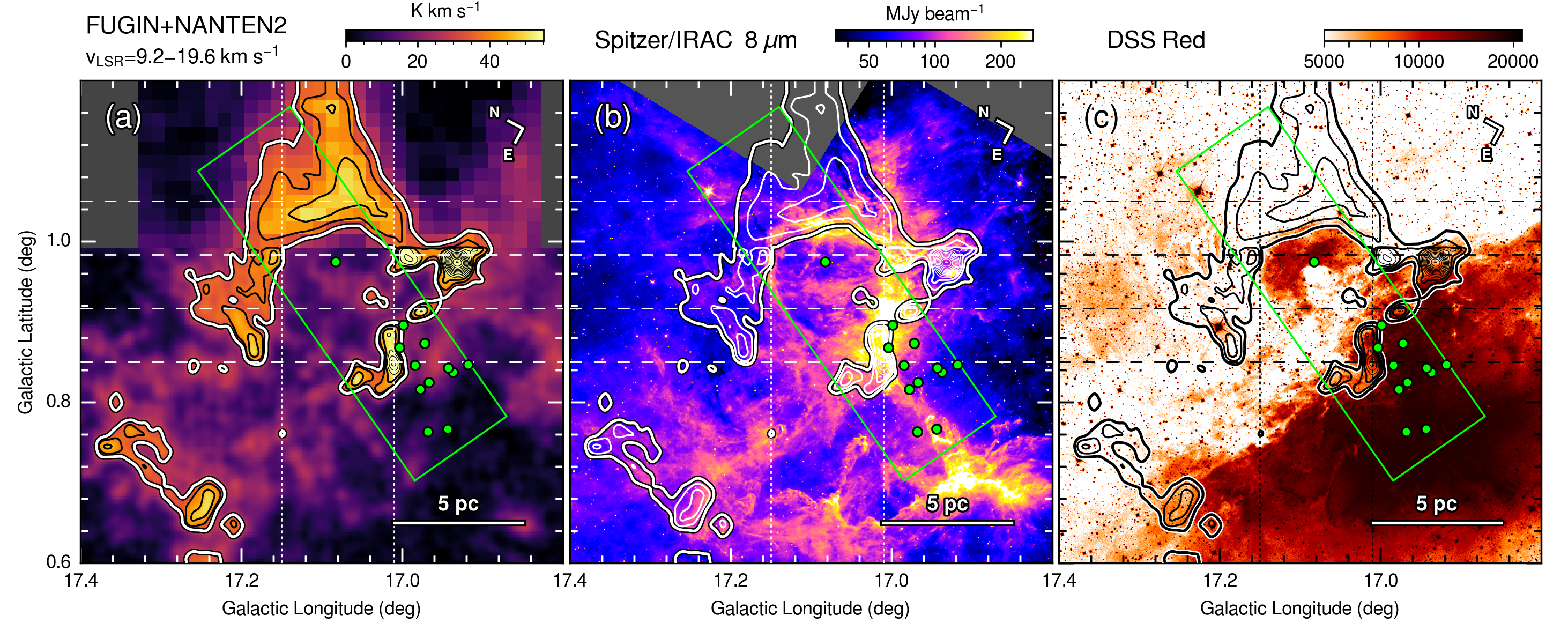}
 \end{center}
\caption{
(a) Spatial distribution of the blue-shifted component of the N cloud.
The velocity range of 9.2--19.6 \kms \ is integrated for \COa \ \Jeq 1--0 emission.
Contour levels are every 8 \Kkms \ from 30 \Kkms.
Dashed horizontal lines are areas where integrated to make longitude-velocity diagrams in Figure \ref{fig:7}.
A green rectangle indicates an area where integrated to make a position-velocity diagram of Figure \ref{fig:9}.
Dotted vertical lines indicate spatial extent of Spitzer bubble N19.
O-type stars \citep{eva05} are shown as green circles.
(b) Map of the 8 $\mu$m emission overlaid with the contours of \COa \ \Jeq 1--0 same as shown in panel a.
(c) Map of the DSS Red image overlaid with the contours of \COa \ \Jeq 1--0 same as shown in panel a.
}
\label{fig:5}
\end{figure}

\begin{figure}[t]
 \begin{center}
  \includegraphics[width=\textwidth]{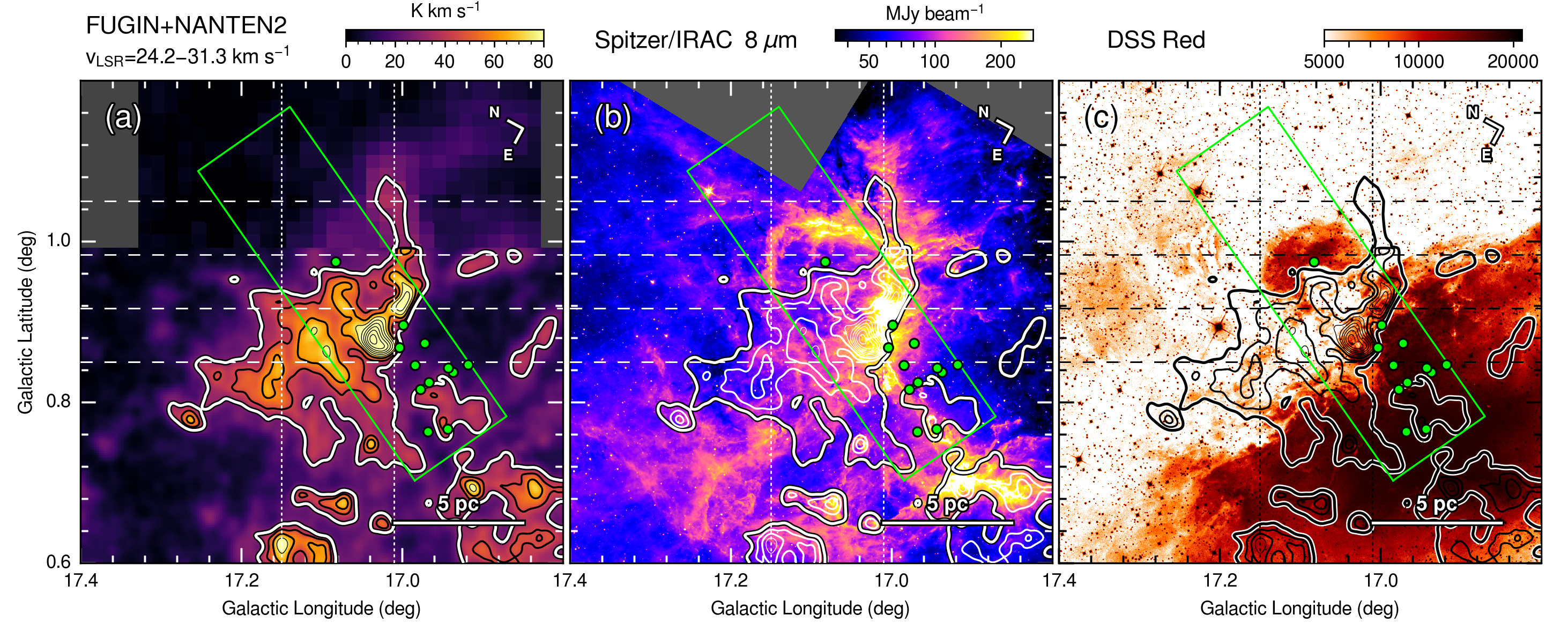}
 \end{center}
\caption{
(a) Spatial distribution of the red-shifted component of the N cloud.
The velocity range of 24.2--31.3 \kms \ is integrated for \COa \ \Jeq 1--0 emission.
Contour levels are every 15 \Kkms \ from 30 \Kkms.
Dashed horizontal lines are areas where integrated to make longitude-velocity diagrams in Figure \ref{fig:7}.
A green rectangle indicates an area where integrated to make a position-velocity diagram of Figure \ref{fig:9}.
Dotted vertical lines indicate spatial extent of Spitzer bubble N19.
O-type stars \citep{eva05} are shown as green circles.
(b) Map of the 8 $\mu$m emission overlaid with the contours of \COa \ \Jeq 1--0 same as shown in panel a.
(c) Map of the DSS Red image overlaid with the contours of \COa \ \Jeq 1--0 same as shown in panel a.
}
\label{fig:6}
\end{figure}

\begin{figure}[t]
 \begin{center}
  \includegraphics[width=0.7\textwidth]{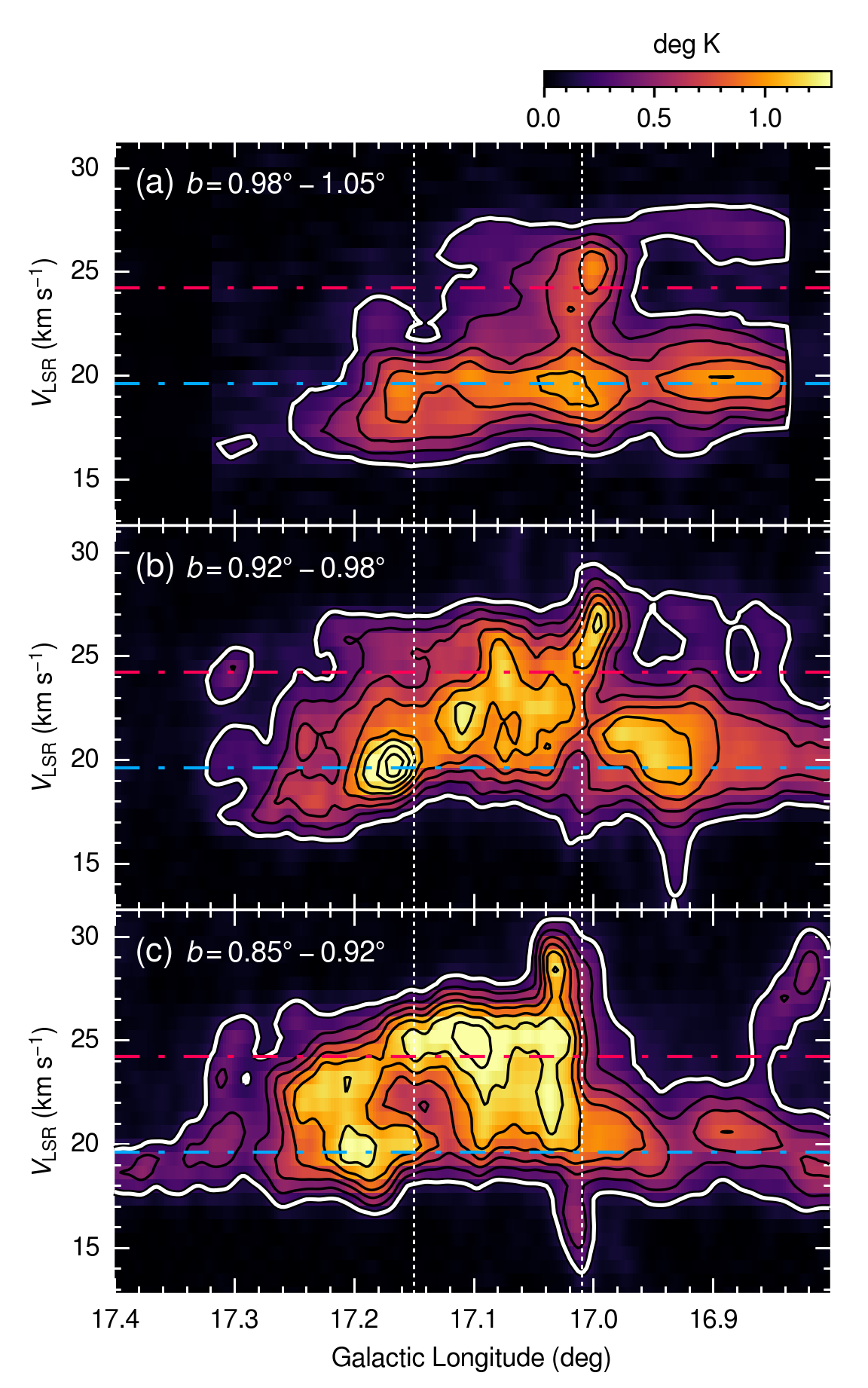}
 \end{center}
\caption{
Longitude-velocity ($l$-$v$) diagrams around the Spitzer bubble N19 for the data of \COa \ \Jeq 1--0 taken with the FUGIN and the NANTEN2 telescope.
The integrated ranges are (a) 0.98\degree--1.05\degree, (b) 0.92\degree--0.98\degree, and (c) 0.85\degree--0.92\degree.
The boundaries of the Spitzer bubble N19 are indicated as dotted vertical lines.
Dash-dotted lines indicate the velocity ranges of the blue- and red-shifted components.
Contour levels are every 0.2 deg K from 0.2 deg K.
}
\label{fig:7}
\end{figure}

\begin{figure}[t]
 \begin{center}
  \includegraphics[width=0.8\textwidth]{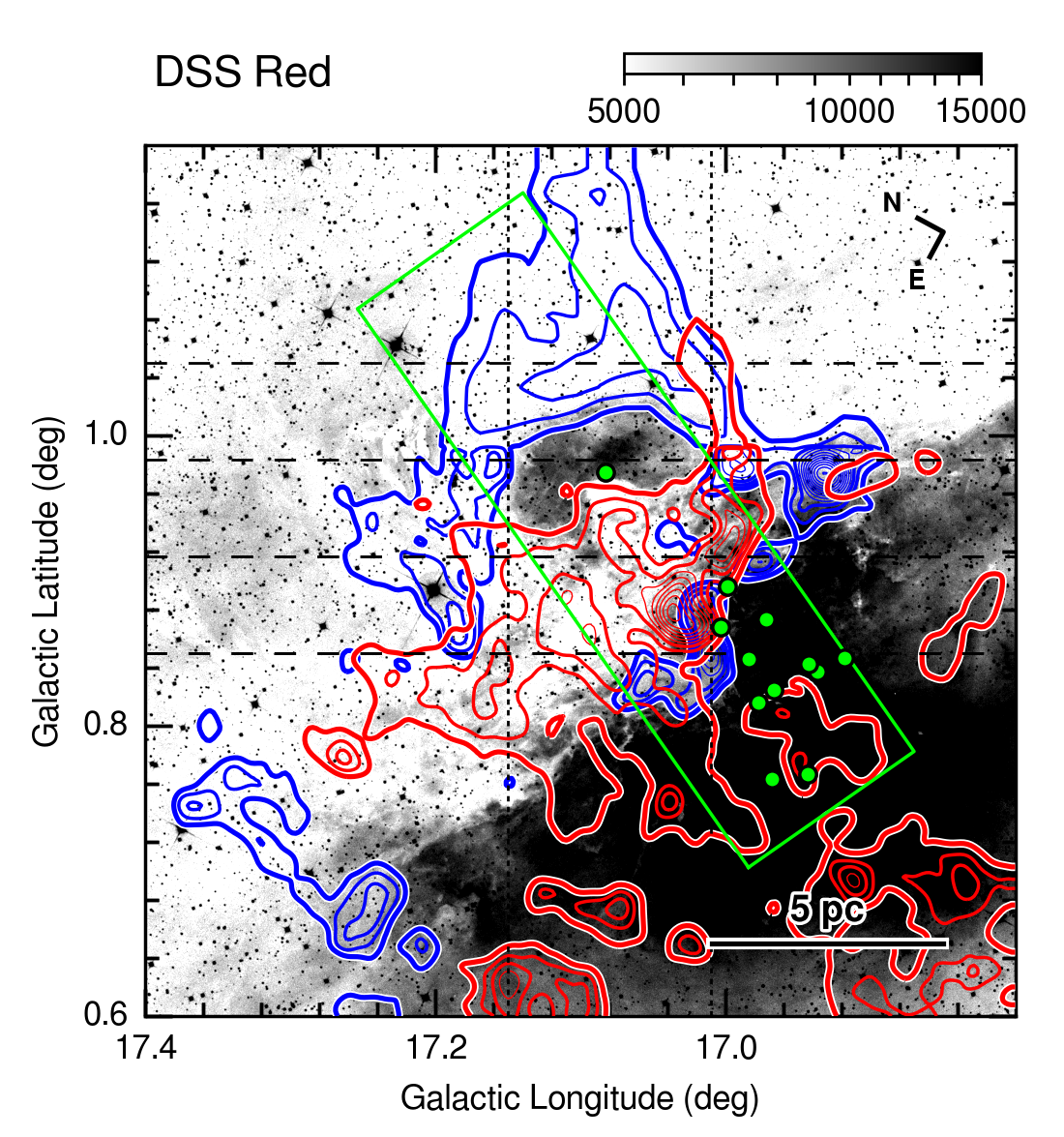}
 \end{center}
\caption{
Complementary distributions of the two velocity components in the N cloud.
Blue contours indicating the blue-shifted component with contour levels of every 8 \Kkms \ from 30 \Kkms \ and Red contours indicating the red-shifted component with contour levels of every 15 \Kkms \ from 30 \Kkms \ are overlaid on a map of the DSS Red image.
Markers and lines are same as Figure \ref{fig:5}.
}
\label{fig:8}
\end{figure}

\begin{figure}[t]
 \begin{center}
  \includegraphics[width=0.8\textwidth]{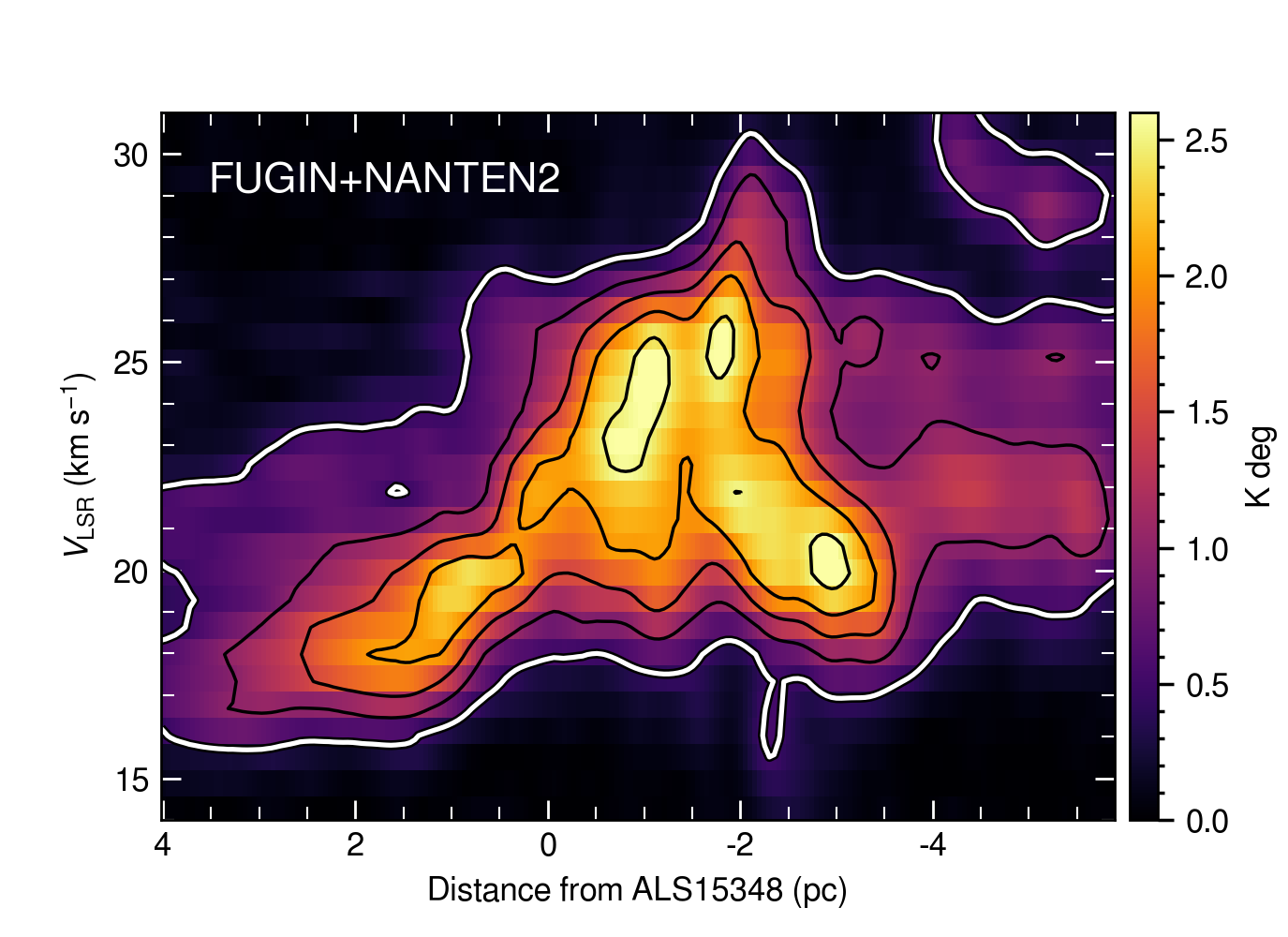}
 \end{center}
\caption{
Position-velocity diagram of \COa \ \Jeq 1--0 toward the area shown as the green rectangle in Figure \ref{fig:5}.
The x axis is a distance from the O star ALS15348, corresponding to the long side of the rectangle, and the emissions are integrated along the short side of the rectangle.
Contour levels are every 0.5 deg K from 0.5 deg K.
}
\label{fig:9}
\end{figure}

\begin{figure}[t]
 \begin{center}
  \includegraphics[width=0.8\textwidth]{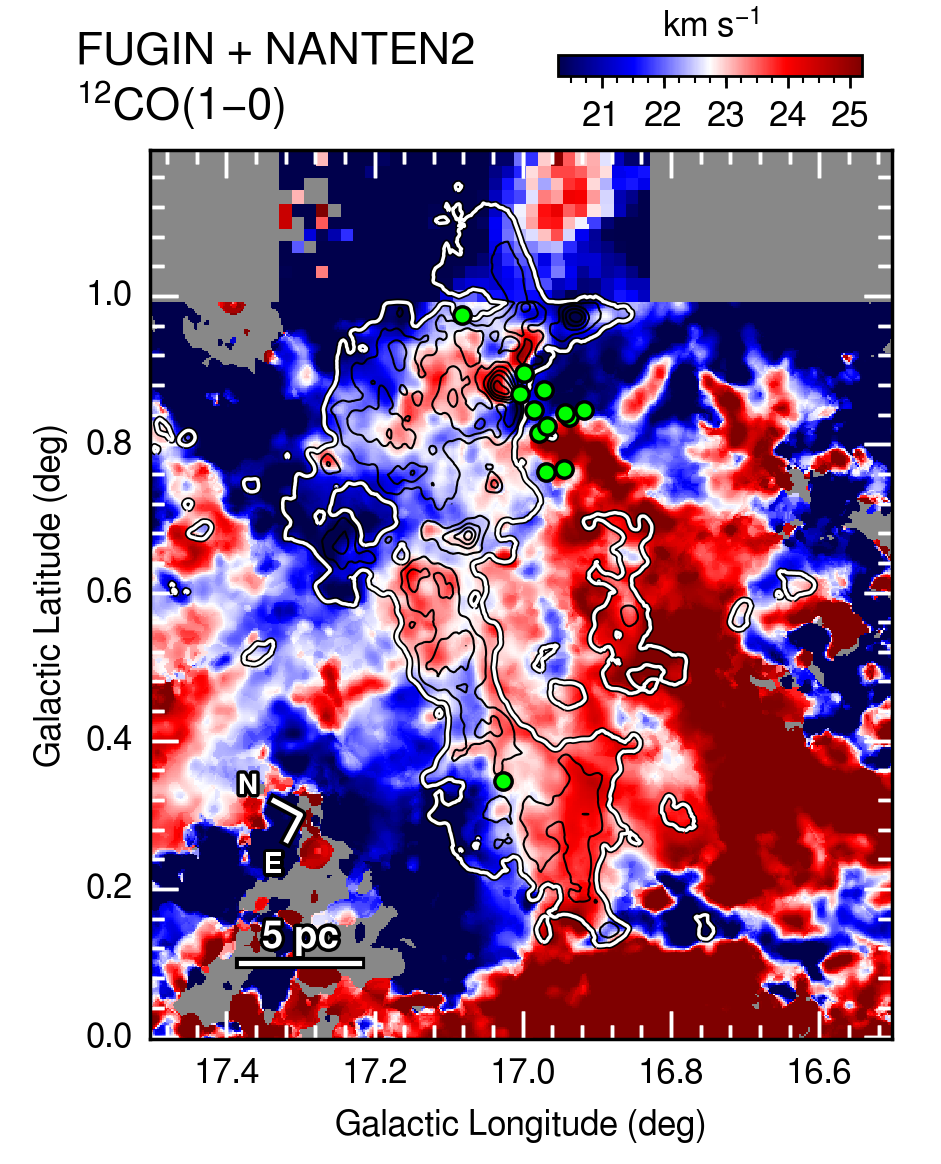}
 \end{center}
\caption{
Intensity-weighted mean velocity maps for \COa \ \Jeq 1--0 data obtained by FUGIN and NANTEN2.
Contours indicate integrated intensity of \COa \ \Jeq 1--0 with contour levels for every 30 \Kkms \ from 80 \Kkms.
O-type stars \citep{eva05} are shown as green circles.
}
\label{fig:10}
\end{figure}

\begin{figure}[t]
 \begin{center}
  \includegraphics[width=0.8\textwidth]{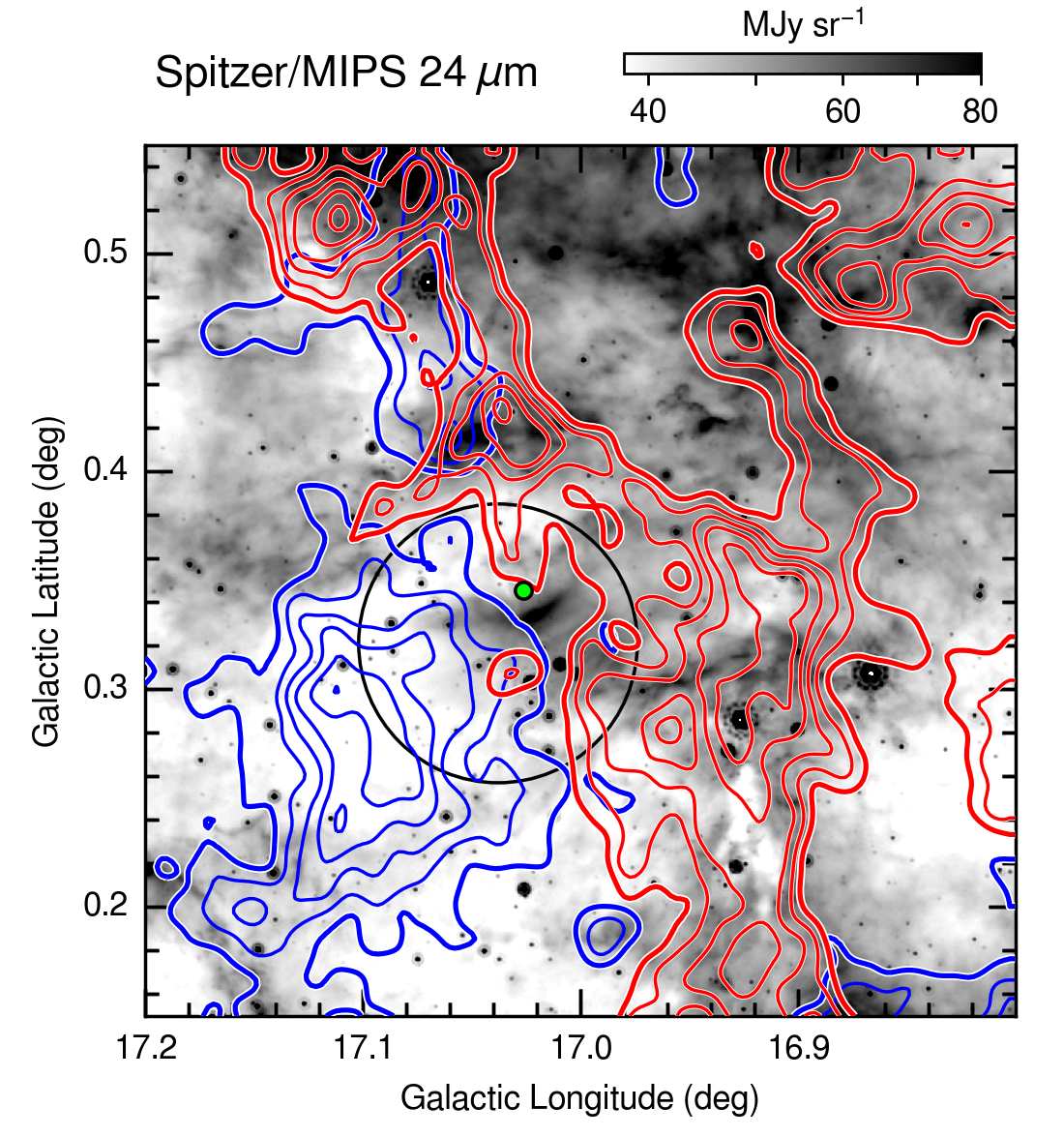}
 \end{center}
\caption{
Complementary distributions of the two velocity components in the SE cloud.
Blue contours indicating the blue-shifted component with contour levels of every 8 \Kkms \ from 13.4 \Kkms \ and Red contours indicating the red-shifted component with contour levels of every 8 \Kkms \ from 33.9 \Kkms \ are overlaid on a map of the 24\ $\mu$m emission.
O-type stars \citep{eva05} are shown as green circles.
Black open circle indicates the position and size of the \HII \ region G017.037+00.320 \citep{and14}.
}
\label{fig:10-1}
\end{figure}

\begin{figure}[t]
 \begin{center}
  \includegraphics[width=0.8\textwidth]{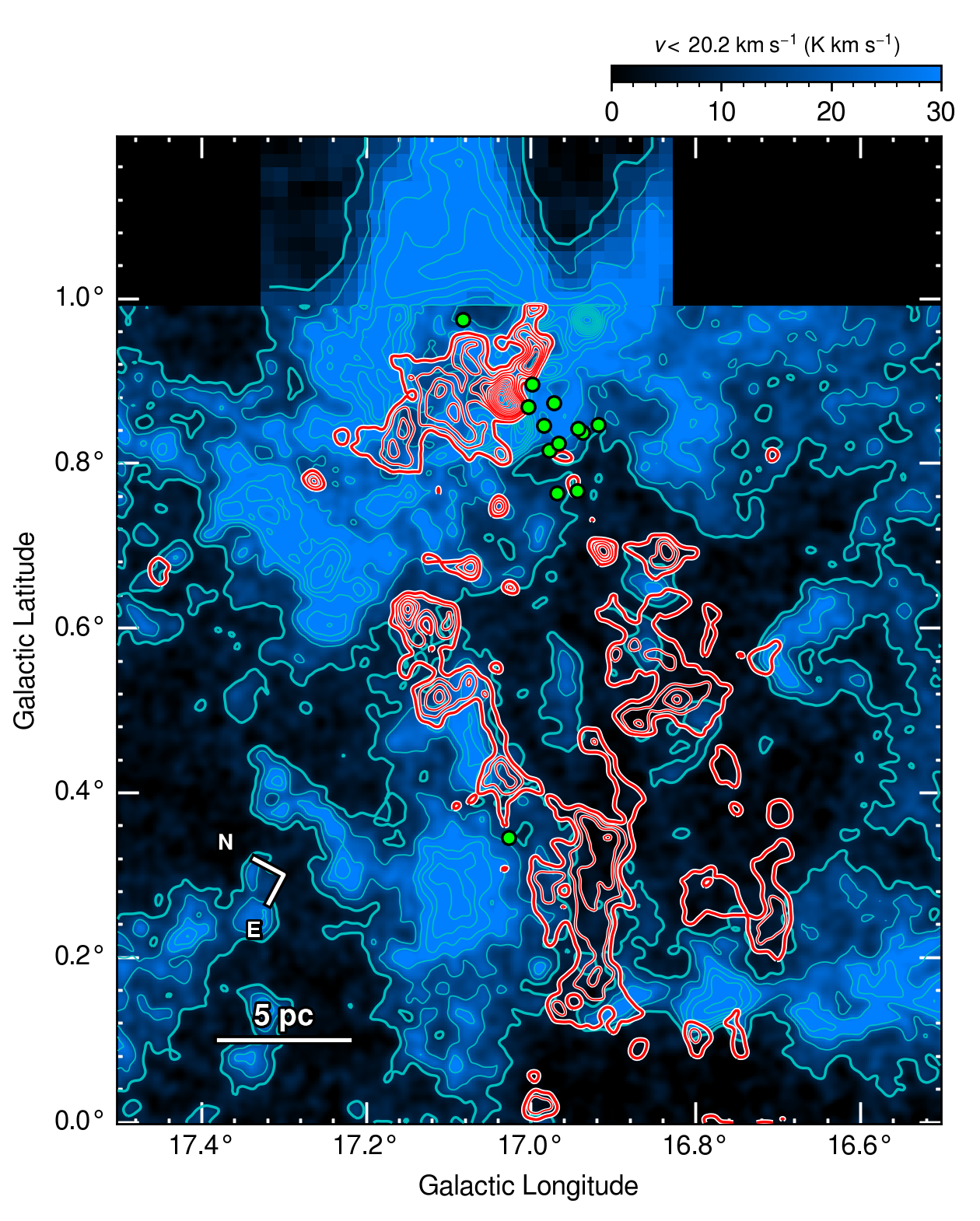}
 \end{center}
\caption{
Image of blue-shifted component with contours of blue- and red-shifted components. 
Velocity integration ranges of blue- and red-shifted components are 9.2 \kms $<$ \vlsr $<$ 20.2 \kms \ and 24.2 \kms $<$ \vlsr $<$ 31.3 \kms, respectively. 
Contour levels for blue- and red-shifted components are every 9 \Kkms \ from 9 \Kkms \ and  10 \Kkms \ from 42 \Kkms, respectively.
O-type stars \citep{eva05} are shown as green circles.
}
\label{fig:11}
\end{figure}

\begin{figure}[t]
 \begin{center}
  \includegraphics[width=\textwidth]{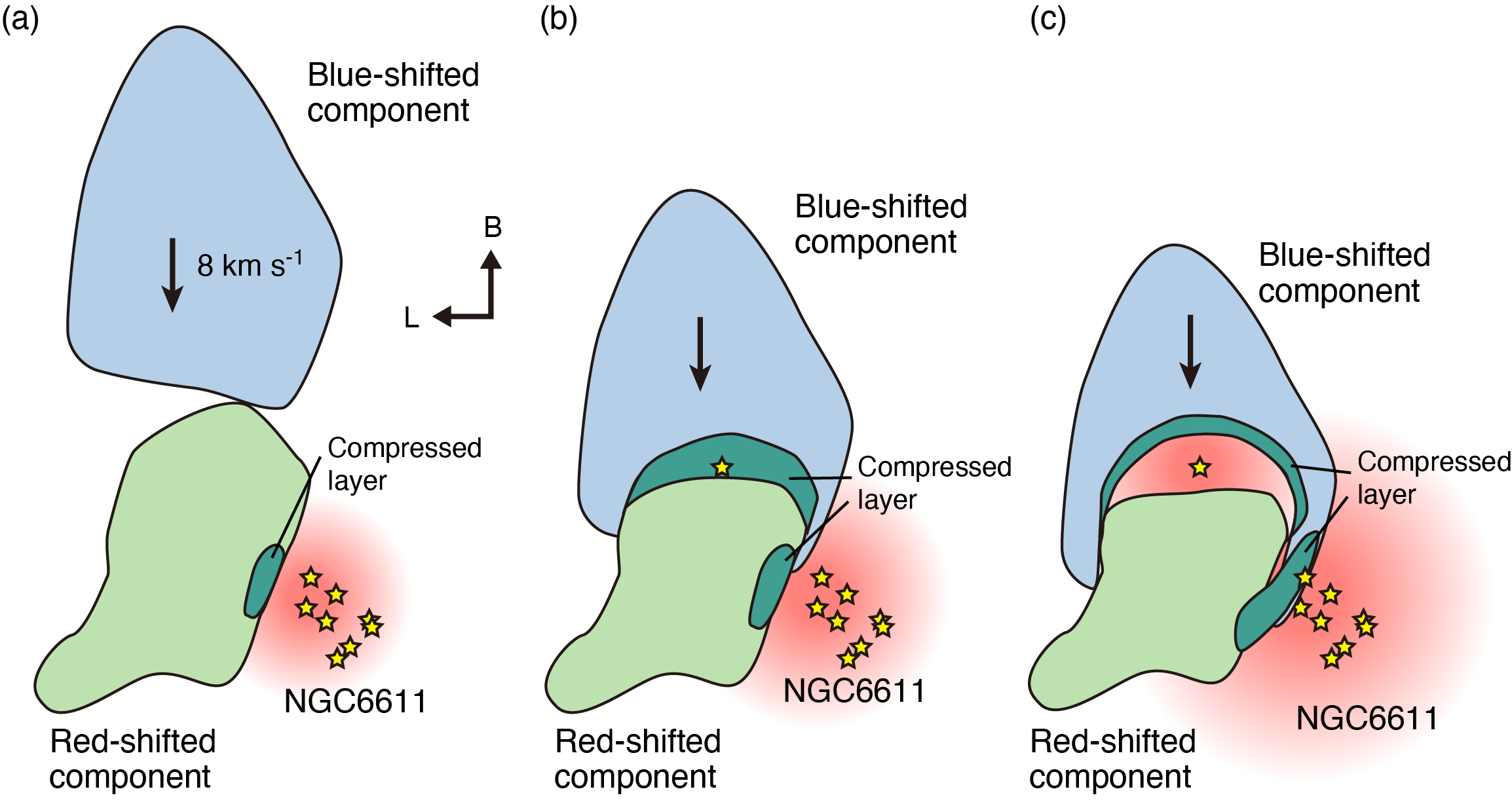}
 \end{center}
\caption{
Schematic diagrams of the collision between the two velocity components of the N cloud seen in the sky.
(a) The two component begin collision $4 \times 10^5$ yr ago.
We assume that the NGC6611 cluster is formed and associated with the red-shifted component at this time.
Boundary of the red-shifted component toward the cluster is interacted with the feedback from the cluster and compressed.
(b) The small cloud (the red-shifted component) created a cavity in the large cloud (the blue-shifted component), and the O star ALS15348 was formed $\sim 10^5$ yr after the collision started.
(c) Current status of the N cloud. 
The cavity was ionized and the small cloud inside the cavity was dissipated by collision and ionization.
Since the ionization is still ongoing, the remnant of the small cloud located behind the ionization front.
The large cloud is colliding from far side of the LOS, therefore the small cloud is located near side to the observer.
}
\label{fig:12}
\end{figure}

\begin{figure}[t]
 \begin{center}
  \includegraphics[width=\textwidth]{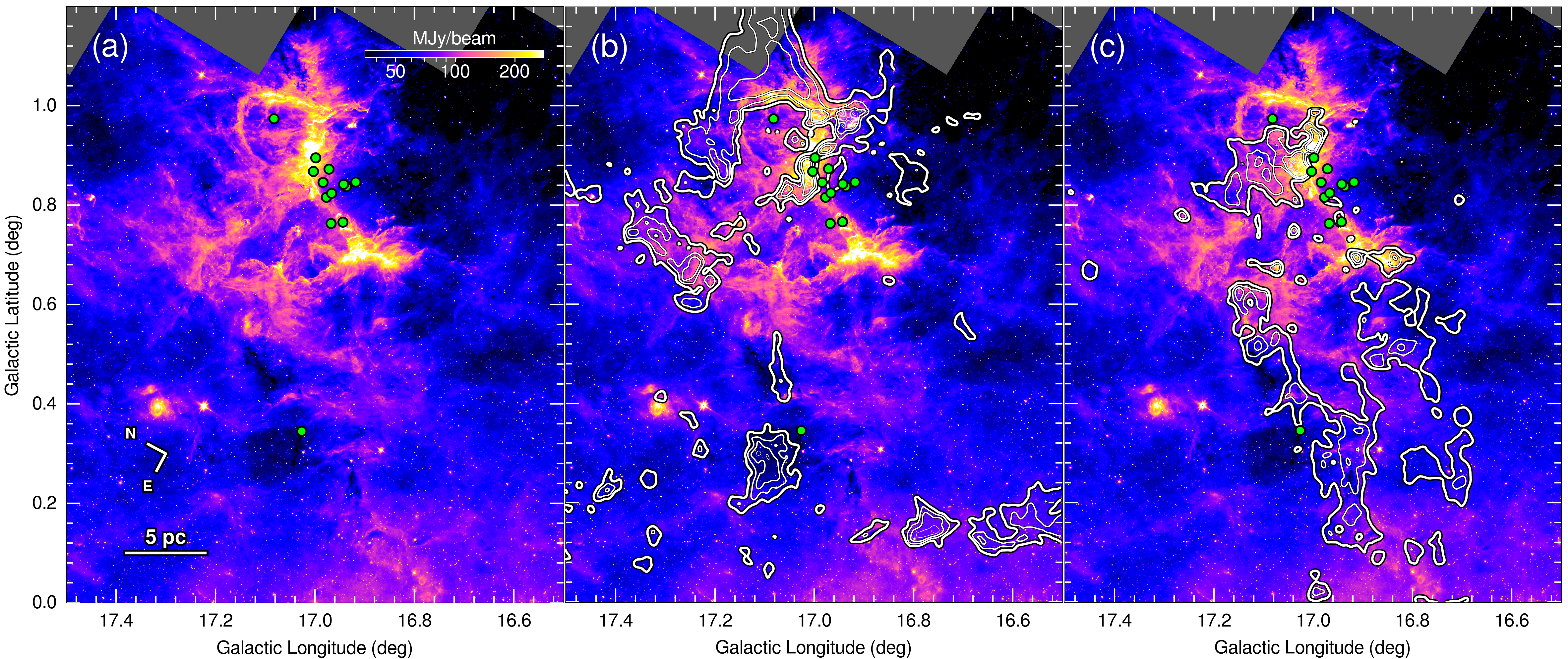}
 \end{center}
\caption{
(a) Map of the 8 $\mu$m emission.
O-type stars \citep{eva05} are shown as green circles.
(b) Same as (a) but overlaid with contours indicating the blue-shifted components.
Contour levels are every 10 \Kkms \ from 22 \Kkms.
(c) Same as (a) but overlaid with contours indicating the red-shifted components.
Contour levels are every 15 \Kkms \ from 40 \Kkms.
}
\label{fig:13}
\end{figure}

\begin{figure}[t]
 \begin{center}
  \includegraphics[width=0.8\textwidth]{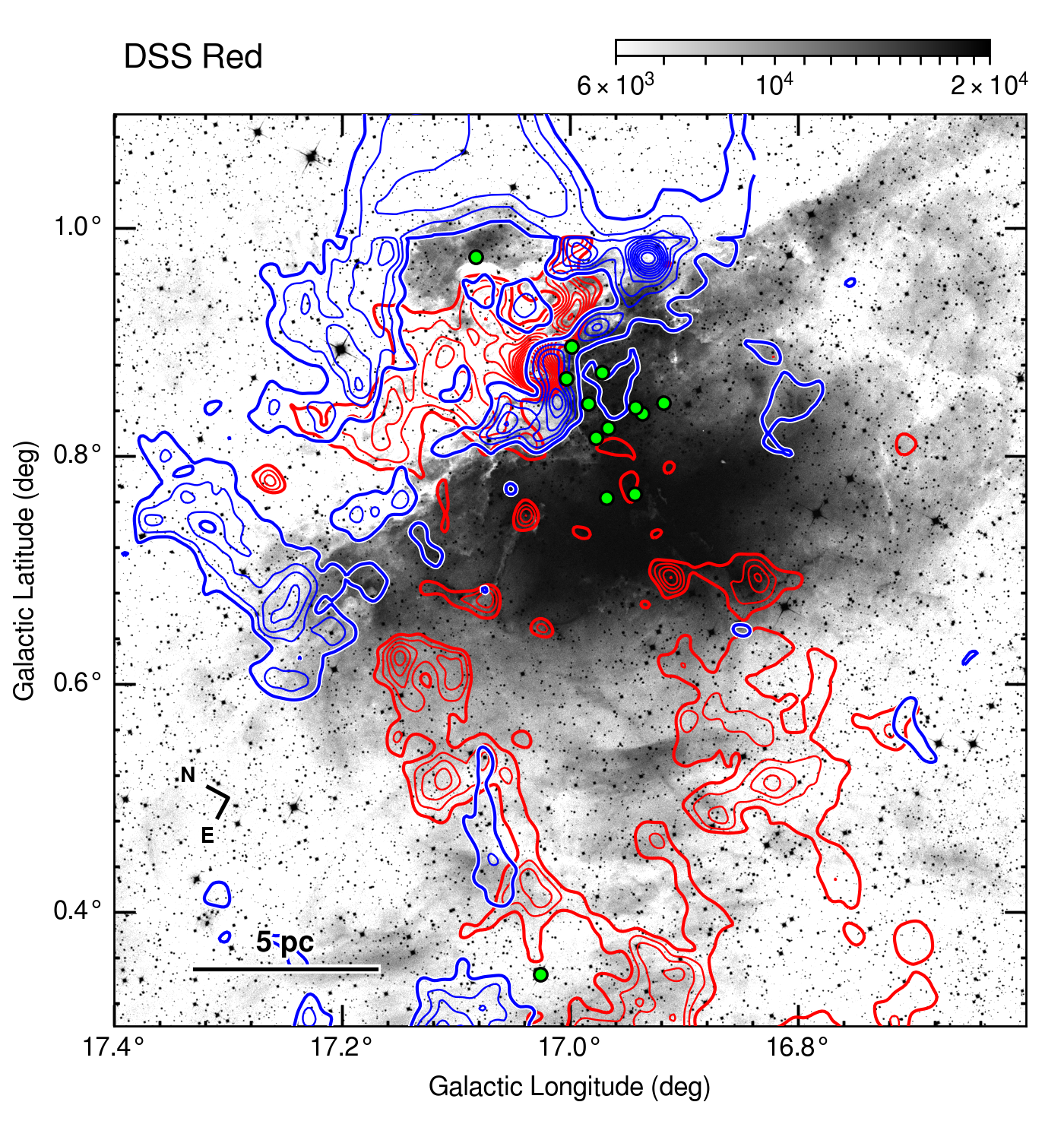}
 \end{center}
\caption{
Complementary distributions of the two velocity components in the M16 GMC around the NGC6611 cluster.
Blue contours indicating the blue-shifted component with contour levels of every 10 \Kkms \ from 30 \Kkms \ and Red contours indicating the red-shifted component with contour levels of every 10 \Kkms \ from 40 \Kkms \ are overlaid on a map of the DSS Red image.
Velocity integration ranges of blue- and red-shifted components are 9.2 \kms $<$ \vlsr $<$ 20.2 \kms \ and 24.2 \kms $<$ \vlsr $<$ 31.3 \kms, respectively.
Markers are same as Figure \ref{fig:5}.
}
\label{fig:15}
\end{figure}

%% file: main.bbl
\begin{thebibliography}{}

\bibitem[Anathpindika(2009a)]{ana09a} Anathpindika, S.\ 2009a, \aap, 504, 437 
\bibitem[Anathpindika(2009b)]{ana09b} Anathpindika, S.\ 2009b, \aap, 504, 451 
\bibitem[Anathpindika(2010)]{ana10} Anathpindika, S.~V.\ 2010, \mnras, 405, 1431
\bibitem[Anderson et al.(2014)]{and14} Anderson, L.~D., Bania, T.~M., Balser, D.~S., Cunningham, V., Wenger, T.~V., Johnstone, B.~M., \& Armentrout, W.~P.\ 2014, \apjs, 212, 1 
\bibitem[Astropy Collaboration et al.(2013)]{ast13} Astropy Collaboration, Robitaille, T.~P., Tollerud, E.~J., et al.\ 2013, \aap, 558, A33 
\bibitem[Belikov et al.(2000)]{bel00} Belikov, A.~N., Kharchenko, N.~V., Piskunov, A.~E., \& Schilbach, E.\ 2000, \aap, 358, 886 
\bibitem[Bisbas et al.(2017)]{bis17} Bisbas, T.~G., Tanaka, K.~E.~I., Tan, J.~C., Wu, B., \& Nakamura, F.\ 2017, \apj, 850, 23 
\bibitem[Bonatto et al.(2006)]{bon06} Bonatto, C., Santos, J.~F.~C., Jr., \& Bica, E.\ 2006, \aap, 445, 567 
\bibitem[Bosch et al.(1999)]{bos99} Bosch, G.~L., Morrell, N.~I., \& Niemel{\"a}, V.~S.\ 1999, Revista Mexicana de Astronomia y Astrofisica, 35, 85 
\bibitem[Churchwell et al.(2006)]{chu06} Churchwell, E., Povich, M.~S., Allen, D., et al.\ 2006, \apj, 649, 759 
\bibitem[Comer{\'o}n \& Torra(2018)]{com18} Comer{\'o}n, F., \& Torra, J. 2018, \aap, 618, A67
\bibitem[Dame et al.(2001)]{dam01} Dame, T.~M., Hartmann, D., \& Thaddeus, P.\ 2001, \apj, 547, 792 
\bibitem[Deharveng et al.(2010)]{deh10} Deharveng, L., et al. 2010, \aap, 523, A6
\bibitem[Dobashi et al.(2014)]{dob14} Dobashi, K., Matsumoto, T., Shimoikura, T., et al.\ 2014, \apj, 797, 58 
\bibitem[Dobashi et al.(2019)]{dob19} Dobashi, K., Shimoikura, T., Katakura, S., Nakamura, F., \& Shimajiri, Y. 2019, \pasj, 71, S12
\bibitem[Dufton et al.(2006)]{duf06} Dufton, P.~L., Smartt, S.~J., Lee, J.~K., et al.\ 2006, \aap, 457, 265 
\bibitem[Elmegreen \& Lada(1977)]{elm77} Elmegreen, B.~G., \& Lada, C.~J.\ 1977, \apj, 214, 725 
\bibitem[Enokiya et al.(2018)]{eno18} Enokiya, R., et al.\ 2018, \pasj, 70, S49 
\bibitem[Enokiya et al.(2019a)]{eno19a} Enokiya, R., Torii, K., \& Fukui, Y. 2019a, \pasj, 127 
\bibitem[Enokiya et al.(2019b)]{eno19b} Enokiya, R., et al. 2019b, arXiv e-prints, arXiv:1912.11607
\bibitem[Evans et al.(2005)]{eva05} Evans, C.~J., Smartt, S.~J., Lee, J.-K., et al.\ 2005, \aap, 437, 467 
\bibitem[Fujita et al.(2017)]{fuj17} Fujita, S., et al. 2017, arXiv e-prints, arXiv:1706.05664
\bibitem[Fujita et al.(2019a)]{fuj19a} Fujita, S., et al. 2019a, \apj, 872, 49 
\bibitem[Fujita et al.(2019b)]{fuj19b} Fujita, S., et al. 2019b, \pasj, 46 
\bibitem[Fukui et al.(2001)]{fuk01} Fukui, Y., Mizuno, N., Yamaguchi, R., Mizuno, A., \& Onishi, T.\ 2001, \pasj, 53, L41 
\bibitem[Fukui et al.(2014)]{fuk14} Fukui, Y., Ohama, A., Hanaoka, N., et al.\ 2014, \apj, 780, 36 
\bibitem[Fukui et al.(2015)]{fuk15} Fukui, Y., Harada, R., Tokuda, K., et al.\ 2015, \apjl, 807, L4 
\bibitem[Fukui et al.(2016)]{fuk16} Fukui, Y., Torii, K., Ohama, A., et al.\ 2016, \apj, 820, 26 
\bibitem[Fukui et al.(2017)]{fuk17a} Fukui, Y., Tsuge, K., Sano, H., Bekki, K., Yozin, C., Tachihara, K., \& Inoue, T.\ 2017, \pasj, 69, L5 
\bibitem[Fukui et al.(2018a)]{fuk17b} Fukui, Y., et al.\ 2018a, \apj, 859, 166  
\bibitem[Fukui et al.(2018b)]{fuk18a} Fukui, Y., et al.\ 2018b, \pasj, 70, S41   
\bibitem[Fukui et al.(2018c)]{fuk18b} Fukui, Y., et al.\ 2018c, \pasj, 70, S44   
\bibitem[Fukui et al.(2018d)]{fuk18c} Fukui, Y., et al.\ 2018d, \pasj, 70, S46   
\bibitem[Fukui et al.(2020)]{fuk20} Fukui, Y., et al.\ 2020, \pasj, accepted   
\bibitem[Furukawa et al.(2009)]{fur09} Furukawa, N., Dawson, J.~R., Ohama, A., et al.\ 2009, \apjl, 696, L115 
\bibitem[Georgelin \& Georgelin(1970)]{geo70} Georgelin, Y.~P., \& Georgelin, Y.~M.\ 1970, \aap, 6, 349 
\bibitem[Guarcello et al.(2007)]{gua07} Guarcello, M.~G., Prisinzano, L., Micela, G., et al.\ 2007, \aap, 462, 245 
\bibitem[Guarcello et al.(2010)]{gua10} Guarcello, M.~G., Micela, G., Peres, G., Prisinzano, L., \& Sciortino, S.\ 2010, \aap, 521, A61 
\bibitem[Gum(1955)]{gum55} Gum, C.~S.\ 1955, \memras, 67, 155 
\bibitem[Habe \& Ohta(1992)]{hab92} Habe, A., \& Ohta, K.\ 1992, \pasj, 44, 203
\bibitem[Hayashi et al.(2018)]{hay18} Hayashi, K., et al.\ 2018, \pasj, 70, S48 
\bibitem[Haworth et al.(2015a)]{haw15a} Haworth, T.~J., et al.\ 2015, \mnras, 450, 10 
\bibitem[Haworth et al.(2015b)]{haw15b} Haworth, T.~J., Shima, K., Tasker, E.~J., Fukui, Y., Torii, K., Dale, J.~E., Takahira, K., \& Habe, A.\ 2015, \mnras, 454, 1634 
\bibitem[Hester et al.(1996)]{hes96} Hester, J.~J., Scowen, P.~A., Sankrit, R., et al.\ 1996, \aj, 111, 2349 
\bibitem[Hester \& Desch(2005)]{hes05} Hester, J.~J., \& Desch, S.~J.\ 2005, Chondrites and the Protoplanetary Disk, 341, 107 
\bibitem[Hill et al.(2012)]{hil12} Hill, T., et al.\ 2012, \aap, 542, A114 
\bibitem[Hillenbrand et al.(1993)]{hil93} Hillenbrand, L.~A., Massey, P., Strom, S.~E., \& Merrill, K.~M.\ 1993, \aj, 106, 1906 
\bibitem[Hosokawa \& Inutsuka(2005)]{hos05} Hosokawa, T., \& Inutsuka, S.-i.\ 2005, \apj, 623, 917 
\bibitem[Hunter(2007)]{hun07} Hunter J. D., 2007, Comput. Sci. Eng., 9, 90
\bibitem[Indebetouw et al.(2007)]{ind07} Indebetouw, R., Robitaille, T.~P., Whitney, B.~A., et al.\ 2007, \apj, 666, 321 
\bibitem[Inoue \& Fukui(2013)]{ino13} Inoue, T., \& Fukui, Y.\ 2013, \apjl, 774, L31 
\bibitem[Inoue et al.(2018)]{ino18} Inoue, T., Hennebelle, P., Fukui, Y., Matsumoto, T., Iwasaki, K., \& Inutsuka, S.-i.\ 2018, \pasj, 70, S53 
\bibitem[Ishii et al.(2019)]{ish19} Ishii, S., Nakamura, F., Shimajiri, Y., Kawabe, R., Tsukagoshi, T., Dobashi, K., \& Shimoikura, T. 2019, \pasj, 71, S9
\bibitem[Kamazaki et al.(2012)]{kam12} Kamazaki, T., et al. 2012, \pasj, 64, 29
\bibitem[Kobayashi et al.(2018)]{kob18} Kobayashi, M.~I.~N., Kobayashi, H., Inutsuka, S.-i., \& Fukui, Y.\ 2018, \pasj, 70, S59 
\bibitem[Kohno et al.(2018a)]{koh18a} Kohno, M., et al.\ 2018, \pasj, 70, S50 
\bibitem[Kohno et al.(2018b)]{koh18b} Kohno, M., et al. 2018b, \pasj, 126 
\bibitem[Krumholz \& McKee(2008)]{kru08} Krumholz, M.~R., \& McKee, C.~F.\ 2008, \nat, 451, 1082 
\bibitem[Kuno et al.(2011)]{kun11} Kuno, N., et al. 2011, in Proc. XXXth URSI General Assembly and Scientific Symposium (New York: IEEE)
\bibitem[Kuwahara et al.(2019)]{kuw19} Kuwahara, S., Torii, K., Mizuno, N., Fujita, S., Kohno, M., \& Fukui, Y. 2019, arXiv e-prints, arXiv:1912.00441
\bibitem[Lada \& Lada(2003)]{lad03} Lada, C.~J., \& Lada, E.~A.\ 2003, \araa, 41, 57 
\bibitem[Levay et al.(2015)]{lev15} Levay, Z.~G., Christian, C.~A., Mack, J., et al.\ 2015, American Astronomical Society Meeting Abstracts, 225, 436.02 
\bibitem[Li et al.(2018)]{li18} Li, Q., Tan, J.~C., Christie, D., Bisbas, T.~G., \& Wu, B.\ 2018, \pasj, 70, S56 
\bibitem[Nakamura et al.(2019)]{nak19} Nakamura, F., et al. 2019, \pasj, 71, S10
\bibitem[Nishimura et al.(2015)]{nis15} Nishimura, A., Tokuda, K., Kimura, K., et al.\ 2015, \apjs, 216, 18 
\bibitem[Nishimura et al.(2018)]{nis18} Nishimura, A., et al.\ 2018, \pasj, 70, S42 
\bibitem[Matsumoto et al.(2015)]{mat15} Matsumoto, T., Dobashi, K., \& Shimoikura, T.\ 2015, \apj, 801, 77 
\bibitem[Matsunaga et al.(2001)]{mat01} Matsunaga, K., Mizuno, N., Moriguchi, Y., et al.\ 2001, \pasj, 53, 1003 
\bibitem[McCaughrean \& Andersen(2002)]{mcc02} McCaughrean, M.~J., \& Andersen, M.\ 2002, \aap, 389, 513 
\bibitem[McClure-Griffiths et al.(2005)]{mcc05} McClure-Griffiths, N.~M., Dickey, J.~M., Gaensler, B.~M., et al.\ 2005, \apjs, 158, 178 
\bibitem[McLeod et al.(2015)]{mcl15} McLeod, A.~F., Dale, J.~E., Ginsburg, A., et al.\ 2015, \mnras, 450, 1057 
\bibitem[Minamidani et al.(2016)]{min16} Minamidani, T., et al. 2016, \procspie, 9914, 99141Z
\bibitem[McMillan et al.(2007)]{mcm07} McMillan, S.~L.~W., Vesperini, E., \& Portegies Zwart, S.~F.\ 2007, \apjl, 655, L45 
\bibitem[Mizuno \& Fukui(2004)]{miz04} Mizuno, A., \& Fukui, Y.\ 2004, Milky Way Surveys: The Structure and Evolution of our Galaxy, 317, 59 
\bibitem[Moriguchi et al.(2002)]{mor02} Moriguchi, Y., Onishi, T., Mizuno, A., \& Fukui, Y.\ 2002, 8th Asian-Pacific Regional Meeting, Volume II, 173 
\bibitem[Ochsenbein et al.(2000)]{och00} Ochsenbein, F., Bauer, P., \& Marcout, J.\ 2000, \aaps, 143, 23 
\bibitem[Ohama et al.(2010)]{oha10} Ohama, A., Dawson, J.~R., Furukawa, N., et al.\ 2010, \apj, 709, 975 
\bibitem[Ohama et al.(2017)]{oha17} Ohama, A., et al. 2017, arXiv e-prints, arXiv:1706.05652
\bibitem[Ohama et al.(2018a)]{oha18a} Ohama, A., et al.\ 2018a, \pasj, 70, S45   
\bibitem[Ohama et al.(2018b)]{oha18b} Ohama, A., et al.\ 2018b, \pasj, 70, S47    
\bibitem[Okamoto et al.(2017)]{oka17} Okamoto, R., Yamamoto, H., Tachihara, K., et al.\ 2017, \apj, 838, 132 
\bibitem[Oliveira(2008)]{oli08} Oliveira, J.~M.\ 2008, Handbook of Star Forming Regions, Volume II, 599 
\bibitem[P{\'e}rez et al.(2007)]{per07} P{\'e}rez F., Granger B., 2007, Comput. Sci. Eng., 9, 21
\bibitem[Pety et al.(2017)]{pet17} Pety, J., et al. 2017, \aap, 599, A98
\bibitem[Pound(1998)]{pou98} Pound, M.~W.\ 1998, \apjl, 493, L113 
\bibitem[Povich et al.(2013)]{pov13} Povich, M.~S., Kuhn, M.~A., Getman, K.~V., et al.\ 2013, \apjs, 209, 31 
\bibitem[Rodgers et al.(1960)]{rod60} Rodgers, A.~W., Campbell, C.~T., \& Whiteoak, J.~B.\ 1960, \mnras, 121, 103 
\bibitem[Sano et al.(2018)]{san18} Sano, H., et al.\ 2018, \pasj, 70, S43 
\bibitem[Sawada et al.(2008)]{saw08} Sawada, T., et al. 2008, \pasj, 60, 445
\bibitem[Sharpless(1959)]{sha59} Sharpless, S.\ 1959, \apjs, 4, 257 
\bibitem[Shima et al.(2018)]{shi18} Shima, K., Tasker, E.~J., Federrath, C., \& Habe, A.\ 2018, \pasj, 70, S54 
\bibitem[Shimoikura et al.(2013)]{shi13} Shimoikura, T., Dobashi, K., Saito, H., et al.\ 2013, \apj, 768, 72 
\bibitem[Shimoikura et al.(2019)]{shi19} Shimoikura, T., Dobashi, K., Hirose, A., Nakamura, F., Shimajiri, Y., \& Sugitani, K. 2019, \pasj, 71, S6
\bibitem[Spitzer(1968)]{spi68} Spitzer, L.\ 1968, New York: Interscience Publication, 1968,  
\bibitem[Sugitani et al.(1986)]{sug86} Sugitani, K., Fukui, Y., Ogawa, H., \& Kawabata, K.\ 1986, \apj, 303, 667 
\bibitem[Sugitani et al.(1991)]{sug91} Sugitani, K., Fukui, Y., \& Ogura, K.\ 1991, \apjs, 77, 59 
\bibitem[Takahira et al.(2014)]{tak14} Takahira, K., Tasker, E.~J., \& Habe, A.\ 2014, \apj, 792, 63 
\bibitem[Takahira et al.(2018)]{tak18} Takahira, K., Shima, K., Habe, A., \& Tasker, E.~J.\ 2018, \pasj, 70, S58 
\bibitem[Takekoshi et al.(2019)]{tak19} Takekoshi, T., et al. 2019, \apj, 883, 156
\bibitem[Tanabe et al.(2019)]{tan19} Tanabe, Y., et al. 2019, /pasj, 71, S8
\bibitem[Torii et al.(2011)]{tor11} Torii, K., Enokiya, R., Sano, H., et al.\ 2011, \apj, 738, 46 
\bibitem[Torii et al.(2015)]{tor15} Torii, K., Hasegawa, K., Hattori, Y., et al.\ 2015, \apj, 806, 7 
\bibitem[Torii et al.(2017)]{tor17} Torii, K., Hattori, Y., Hasegawa, K., et al.\ 2017, \apj, 835, 142 
\bibitem[Torii et al.(2018a)]{tor18a} Torii, K., et al.\ 2018a, \pasj, 70, S51 
\bibitem[Torii et al.(2018b)]{tor18b} Torii, K., et al.\ 2018b, \pasj, 121 
\bibitem[Torii et al.(2019)]{tor19} Torii, K., et al. 2019, \pasj, 71, S2
\bibitem[Umemoto et al.(2017)]{ume17} Umemoto, T., et al. 2017, \pasj, 69, 78
\bibitem[Walker(1961)]{wal61} Walker, M.~F.\ 1961, \apj, 133, 438 
\bibitem[Walt et al.(2011)]{wal11} Walt S. v. d., Colbert S. C., Varoquaux G., 2011, Comput. Sci. Eng., 13, 22
\bibitem[Westerhout(1958)]{wes58} Westerhout, G.\ 1958, \bain, 14, 215 
\bibitem[Whitworth et al.(2018)]{whi18} Whitworth, A., Lomax, O., Balfour, S., M{\`e}ge, P., Zavagno, A., \& Deharveng, L.\ 2018, \pasj, 70, S55 
\bibitem[Wu et al.(2015)]{wu15} Wu, B., Van Loo, S., Tan, J.~C., \& Bruderer, S.\ 2015, \apj, 811, 56 
\bibitem[Wu et al.(2017a)]{wu17a} Wu, B., Tan, J.~C., Nakamura, F., Van Loo, S., Christie, D., \& Collins, D.\ 2017a, \apj, 835, 137 
\bibitem[Wu et al.(2017b)]{wu17b} Wu, B., Tan, J.~C., Christie, D., Nakamura, F., Van Loo, S., \& Collins, D.\ 2017b, \apj, 841, 88 
\bibitem[Wu et al.(2018)]{wu18} Wu, B., Tan, J.~C., Nakamura, F., Christie, D., \& Li, Q.\ 2018, \pasj, 70, S57 
\bibitem[Wolff et al.(2007)]{wol07} Wolff, S.~C., Strom, S.~E., Dror, D., \& Venn, K.\ 2007, \aj, 133, 1092 
\bibitem[Xu et al.(2018)]{xu18} Xu, Y., Bian, S.~B., Reid, M.~J., et al.\ 2018, \aap, 616, L15 
\bibitem[Xu et al.(2019)]{xu19} Xu, J.-L., et al.\ 2019, \aap, 627, A27 
\bibitem[Yamagishi et al.(2016)]{yam16} Yamagishi, M., et al. 2016, \apj, 833, 163
\bibitem[Yamagishi et al.(2018)]{yam18} Yamagishi, M., et al. 2018, \apjs, 235, 9
\bibitem[Yoda et al.(2010)]{yod10} Yoda, T., Handa, T., Kohno, K., et al.\ 2010, \pasj, 62, 1277 
\bibitem[Zinnecker \& Yorke(2007)]{zin07} Zinnecker, H., \& Yorke, H.~W.\ 2007, \araa, 45, 481 

\end{thebibliography}
